\documentclass[sigconf]{acmart}
\usepackage[utf8]{inputenc}
\usepackage{xspace}
\usepackage{makecell}
\usepackage{pifont}
\newcolumntype{L}[1]{>{\raggedright\let\newline\\\arraybackslash\hspace{0pt}}m{#1}}
\newcolumntype{C}[1]{>{\centering\let\newline\\\arraybackslash\hspace{0pt}}m{#1}}
\newcolumntype{R}[1]{>{\raggedleft\let\newline\\\arraybackslash\hspace{0pt}}m{#1}}
\newcommand{\cmark}{\ding{51}}
\newcommand{\xmark}{\ding{55}}

\title{Can Capacitive Touch Images Enhance Mobile Keyboard Decoding?}

\author{Piyawat Lertvittayakumjorn}
\affiliation{
  \institution{Google}
  \city{Mountain View}
  \state{CA}
  \country{USA}}
\email{piyawat@google.com}

\author{Shanqing Cai}
\affiliation{
  \institution{Google}
  \city{Mountain View}
  \state{CA}
  \country{USA}}
\email{cais@google.com}

\author{Billy Dou}
\affiliation{
  \institution{Google}
  \city{Mountain View}
  \state{CA}
  \country{USA}}
\email{billydou@google.com}

\author{Cedric Ho}
\affiliation{
  \institution{Google}
  \city{Mountain View}
  \state{CA}
  \country{USA}}
\email{cedricho@google.com}

\author{Shumin Zhai}
\affiliation{
  \institution{Google}
  \city{Mountain View}
  \state{CA}
  \country{USA}}
\email{zhai@acm.org}

\copyrightyear{2024}
\acmYear{2024}
\setcopyright{rightsretained}
\acmConference[UIST '24]{The 37th Annual ACM Symposium on User Interface Software and Technology}{October 13--16, 2024}{Pittsburgh, PA, USA}
\acmBooktitle{The 37th Annual ACM Symposium on User Interface Software and Technology (UIST '24), October 13--16, 2024, Pittsburgh, PA, USA}
\acmDOI{10.1145/3654777.3676420}
\acmISBN{979-8-4007-0628-8/24/10}

\begin{document}

\newcommand{\cf}{$C$\xspace}
\newcommand{\hf}{$H_f$\xspace}
\newcommand{\hov}{$H_o$\xspace}
\newcommand{\chf}{$CH_f$\xspace}
\newcommand{\chov}{$CH_o$\xspace}

\begin{abstract}

Capacitive touch sensors capture the two-dimensional spatial profile (referred to as a touch heatmap) of a finger's contact with a mobile touchscreen. 
However, the research and design of touchscreen mobile keyboards -- one of the most speed and accuracy demanding touch interfaces -- has focused on the location of the touch centroid derived from the touch image heatmap as the input, discarding the rest of the raw spatial signals. 
In this paper, we investigate whether touch heatmaps can be leveraged to further improve the tap decoding accuracy for mobile touchscreen keyboards. 
Specifically, we developed and evaluated machine-learning models that interpret user taps by using the centroids and/or the heatmaps as their input and studied the contribution of the heatmaps to model performance. 
The results show that adding the heatmap into the input feature set led to 21.4\% relative reduction of character error rates on average, compared to using the centroid alone. 
Furthermore, 
we conducted a live user study with the centroid-based and heatmap-based decoders built into Pixel 6 Pro devices
and observed lower error rate, faster typing speed, and higher self-reported satisfaction score based on the heatmap-based decoder than the centroid-based decoder. 
These findings underline the promise of utilizing touch heatmaps for improving typing experience in mobile keyboards.
\end{abstract}



\maketitle

\section{Introduction} \label{sec:intro}

\begin{figure}[t]
  \centering
  \includegraphics[width=0.9\linewidth]{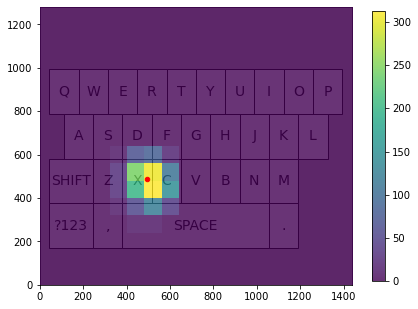}
  \caption{An example of 
  the touch heatmap (the $16 \times 18$ colored grid laying on the keyboard layout) 
  and the derived touch centroid (the red dot).
  In this example, the user aimed to type "c"; however, the derived touch centroid is on the key "x". Applying simple baselines (i.e., the On-key and the Distance baselines) would lead to an incorrect key prediction.
  \textit{Resolution scale}: 1 pixel $\approx$ 0.05 mm.
  } \label{fig:heatmap_example}
  \Description{A figure of size 1440 (width) $\times$ 1280 (height) pixels with a QWERTY keyboard layout in the middle. A heatmap of size $18 \times 16$ is overlaid on top of the figure, showing the area tapped by a user. There is also a touch centroid, in red, plotted on top of the active heatmap area.}
\end{figure}

Tap typing is the most widely used method of text entry on mobile touchscreens.
As smartphone keyboards are relatively small and do not have physical boundaries between keys, the system might interpret a user tap (represented as a \textit{touch point} on the device) differently from what the user intends to type. This leads to a specific category of typing errors, called
\textit{spatial errors}, e.g., \texttt{shock} $\rightarrow$ \texttt{sjock} (\texttt{h} misinterpreted as \texttt{j}) and \texttt{breathing} $\rightarrow$ \texttt{beeathing} (\texttt{r} misinterpreted as \texttt{e}) in the QWERTY keyboard layout.
These errors reduce the user's effective typing speed and hence negatively impact user experience.
Spatial errors are often nicknamed \textit{the fat finger problem}, meaning that users struggle to tap precisely on a touchscreen because 
human fingers are relatively large and covered with soft skin, producing hard-to-control contact areas on obscured target keys \cite{vogel2007shift}.
An alternative explanation to this phenomenon is \textit{the perceived input point model}, stating that the center of the touch area, as reported by the device, usually locates at an offset below the user's aimed position and this offset varies depending on various factors such as the user's hand posture and their typing mental model \cite{holz2010generalized}.
According to the Finger-Fitts Law model (FFitts Law) \cite{bi2013ffitts}, the variability of the touch centroid positions could originate from not only the speed-accuracy tradeoff but also the absolute precision uncertainty of the finger tap action per se.

To deal with this problem, previous studies employed information beyond the touch centroid with the goal of achieving more accurate key decoding.
Such information includes properties of the user tap (e.g., the tap size \cite{goel2013contexttype} and the touch pressure \cite{weir2014uncertain}) as well as contextual information of the typing (e.g., the device motion \cite{goel2012walktype}, the previously typed text \cite{goodman2002language,gunawardana2010usability}, the time elapsed between taps \cite{goel2013contexttype}, and the user identity \cite{sivek2022spatial}).
To our knowledge, a piece of information that has not been used for key decoding in the literature is \textit{the capacitive image} of the user tap, i.e., a two-dimensional spatial data captured by the capacitive touchscreen's contact sensors. 
We also refer to it as a \textit{touch heatmap}, or simply a \textit{heatmap} in this paper.
Fig.~\ref{fig:heatmap_example} shows an example of the touch heatmap together with the touch centroid on a QWERTY keyboard layout.
In practice, the centroid is derived from the heatmap and used as a summary of the touch pattern, discarding some spatial information that may be useful in decoding typing touches.

In this paper, we aim to answer whether the tap-typing decoding accuracy can be improved by leveraging the information carried by touch heatmaps beyond touch centroids.
We trained logistic-regression models\footnote{After experimenting with several machine learning techniques (e.g., logistic regression, neural networks, gradient boosting, and random forest), we chose to conduct experiments on logistic regression because (1) it came out as one of the most accurate models among the machine learning techniques tested (2) it is a simple and lightweight model that can be easily deployed on device and (3) its simplicity allows us to firmly attribute the performance gain to the different input feature sets.} on data collected from
users copy-typing texts on a smartphone with heatmaps logged. The trained models were
evaluated in two stages, the first based on offline datasets
that tested the models' generalizability to unseen users and
phrases and the second based on online deployment of the models
to measure their the practical effectiveness during real-time mobile text entry.

The main contributions of this paper include 
\begin{itemize}
    \item By conducting a modeling study, we demonstrate that the capacitive touch images (i.e., the touch heatmaps) contain information useful for tap-typing decoding and not present in the traditionally-used centroids. 
    The incorporation of touch heatmaps resulted in a 21.4\% relative reduction in character error rate (CER) compared to the centroid-only baseline CER of 4.22\%. With the assistance of language models and additional techniques, the relative CER reduction was 29.7\% from the centroid-only baseline CER of 2.87\%.
    \item We propose a novel representation of touch heatmaps, called \textit{heatmap overlap vector}, which enhances generalizability of the trained spatial model, allowing it to work effectively for inputting unseen text.
    \item We built a heatmap-based spatial model on Pixel 6 Pro devices and measured users’ live typing performance in a lab study. Objective typing metrics and subjective survey results both show advantages of the heatmaps over the centroids for key decoding.
    \item We discuss important considerations for deploying this touch heatmap-based approach in production including generalizability and alternative model architectures.
\end{itemize}

The rest of this paper is organized as follows.
Section~\ref{sec:relatedwork} provides the background and related work on mitigating spatial errors and exploiting heatmap images in user interface research.
Section~\ref{sec:method} describes the spatial models we designed
for decoding key taps from centroids and heatmaps as well as supporting techniques for ensuring smooth user typing experience. 
These formed the basis of a series of three studies in which we systematically evaluated the heatmap-based keyboard decoding models against centroid-only baselines. 
Study 1 (Section~\ref{sec:modelingstudy}) focused on the collection of an initial
dataset that enabled us to train and preliminarily evaluate heatmap-based decoding models. 
Study 2 (Section~\ref{sec:ood}) evaluated the generalizability of the models from Study 1
on new datasets collected based on different sets of text and from new users. 
In Study 3 (Section~\ref{sec:userstudy}), an online user study was conducted to deploy and evaluate the trained spatial models during real-time tap typing.
Section~\ref{sec:discussions} discusses considerations, limitations, and future work, before we conclude the paper in Section~\ref{sec:conclusion}.

\section{Related Work} \label{sec:relatedwork}

\subsection{Mitigating Spatial Errors for Smartphone Text Entry} \label{subsec:spatialerrors}
Researchers have been studying and addressing the spatial errors for mobile keyboards for many years. For example,
\citeauthor{azenkot2012touch} \cite{azenkot2012touch} collected user touch data for different hand postures and discovered that the touch centroids for each user and key can be modeled as a bivariate Gaussian distribution of which the mean has a certain offset from the key center.
Such offsets are different across keys, postures, and participants, so they recommended conducting vertical and horizontal corrections (based on the hand posture used) and personalization for each user.
This is in line with several works 
that emphasize the differences of touch point distributions among users, devices, postures, and directions of finger movement \cite{findlater2011typing,henze2011100,holz2010generalized,kolly2012personal,ma2021modeling}.
ContextType \cite{goel2013contexttype} tried to infer the posture first (using tap sizes and time elapsed between taps) and then employed a posture-specific spatial model to predict the intended key.
WalkType \cite{goel2012walktype} incorporates features derived from the accelerometer in the device to compensate for imprecise input when the user is walking. 
PalmBoard \cite{yi2020palmboard} decoded five-finger typing on a touchpad by extracting features from touch images beyond the centroid coordinates (including duration, area, and pressure).
To the best of our knowledge, there is no existing work leveraging the full information in touch heatmaps to create better spatial models in a data-driven fashion.

Another common approach to improve the text entry performance is to leverage a character-level language model, in addition to the spatial model.
A language model of this sort estimates the probability of the next character based on the previously committed text. For instance, if the user starts with the letter \texttt{s}, the language model can predict that the next letter is more likely to be \texttt{h} than \texttt{j}, helping to prevent spatial errors like the \texttt{shock} $\rightarrow$ \texttt{sjock} example.
Some previous works use the language model scores to adjust the keyboard layout or to add visual clues, highlighting the keys most likely to be tapped next \cite{al2009bigkey,magnien2004mobile}, whereas other works combine the language model scores with the spatial model scores to better predict the user's intended key \cite{goel2013contexttype,goodman2002language,sivek2022spatial}.
In this work, we opt for the latter approach and demonstrate that, after combining with language model scores, touch heatmaps still provide accuracy gains over touch centroids (see Section~\ref{sec:modelingstudy}). 

\subsection{Capacitive Touchscreen}
A capacitive touchscreen is a device's display screen that can capture an image of the finger's contact area at specific points in time. In particular, Projective Capacitive Touch (PCT) \cite{barrett2010projected} is a type of it widely used in portable devices such as smartphones and tablets.
Each PCT sensor is a pair of electrodes separated by a dielectric layer. This structure can act as a capacitor, holding a certain amount of charge, and its capacitance will change when another conductive object (e.g., a finger or a conductive stylus) approaches it.
In a device, PCT sensors are usually arranged as a grid under the screen's cover glass, producing two-dimensional (matrix-like) heatmap images (so called capacitive images).  However, the heatmap image resolution is much lower than the device's display screen resolution (e.g., $39 \times 18$ versus $3120 \times 1440$ for Google Pixel 6 Pro we used in the experiments). 
To effectively use signals from PCT sensors, a touch controller is needed for data pre-processing such as removing noises and deriving touch centroids \cite{wang2011projected}. 
For more details about PCT, we refer the readers to \citeauthor{quinn2021deep} \cite{quinn2021deep}.
Researchers have tried to enhance the commercial touch sensing technology with, for example, frequency variation \cite{sato2012touche} and acoustic sensing \cite{harrison2011tapsense} for novel interactions.
Mass-produced low resolution PCT sensing, however, has the benefit of year-over-year quality and power refinement and integration to enable the usability of basic touchscreen input (e.g., icon level tap, swipe gestures, and multi-finger zoom). 
Sensing subtle gestural differences from today's low resolution PCT sensors is hard, but we have seen recent progress in research and practice due to the advancement of ML technologies and interaction design (e.g., \cite{quinn2021deep}).

Researchers in HCI have investigated heatmap images from capacitive sensors for several UI innovations. 
For example, some have explored ways to extract useful information from these sensor arrays, e.g., obtaining super-resolution images of the touch areas \cite{streli2021capcontact, rusu2023deep} and estimating hand postures \cite{ahuja2021touchpose, choi20213d}, to enable novel interactions on touchscreen devices.
\citeauthor{guarneri2013pca} \cite{guarneri2013pca} developed a pipeline (using Principal Component Analysis \cite{wold1987principal} and the decision tree method) for differentiating one finger, two fingers, and palm, based on the capacitive images.
Similarly, \citeauthor{le2018palmtouch} \cite{le2018palmtouch} proposed PalmTouch, using a Convolution Neural Network (CNN) to classify if a touch is caused by a finger or a palm.
\citeauthor{mayer2017estimating} \cite{mayer2017estimating}, \citeauthor{boceck2019force} \cite{boceck2019force}, and \citeauthor{quinn2021deep} \cite{quinn2021deep} used neural networks to process capacitive images for predicting finger orientation, touch pressure, and touch gestures (e.g., tap, press, or scroll), respectively.

Overall, these recent papers shed light on the benefits of capacitive images via machine learning for many HCI applications. 
Together with the growing on-device computational power, they motivate us to explore the use of capacitive images for touchscreen keyboard decoding.
Due to the extensive optimization and improvement of touchscreen keyboards over the past decades, further gains in decoding are hard to achieve without leveraging new modalities of signals in the user input.
This leads to the interesting question posed in our paper's title: ``Can capacitive touch images enhance mobile keyboard decoding?''.
The rationale for using touch images is that traditional centroid-based decoding assumes point-like taps, but in reality, finger-screen contact areas are not points. Ignoring touch shapes and pressures may lead to incomplete information for decoding ambiguous taps. 
Therefore, in this paper, we preprocess the capacitive images and feed them into logistic regression models for keyboard decoding which, to our knowledge, has not been studied in previous work.

\section{Decoder and Supporting Techniques} \label{sec:method}
\begin{figure}[t]
  \centering
  \includegraphics[trim={1.5cm 0.5cm 1.75cm 0cm},clip,width=0.98\linewidth]{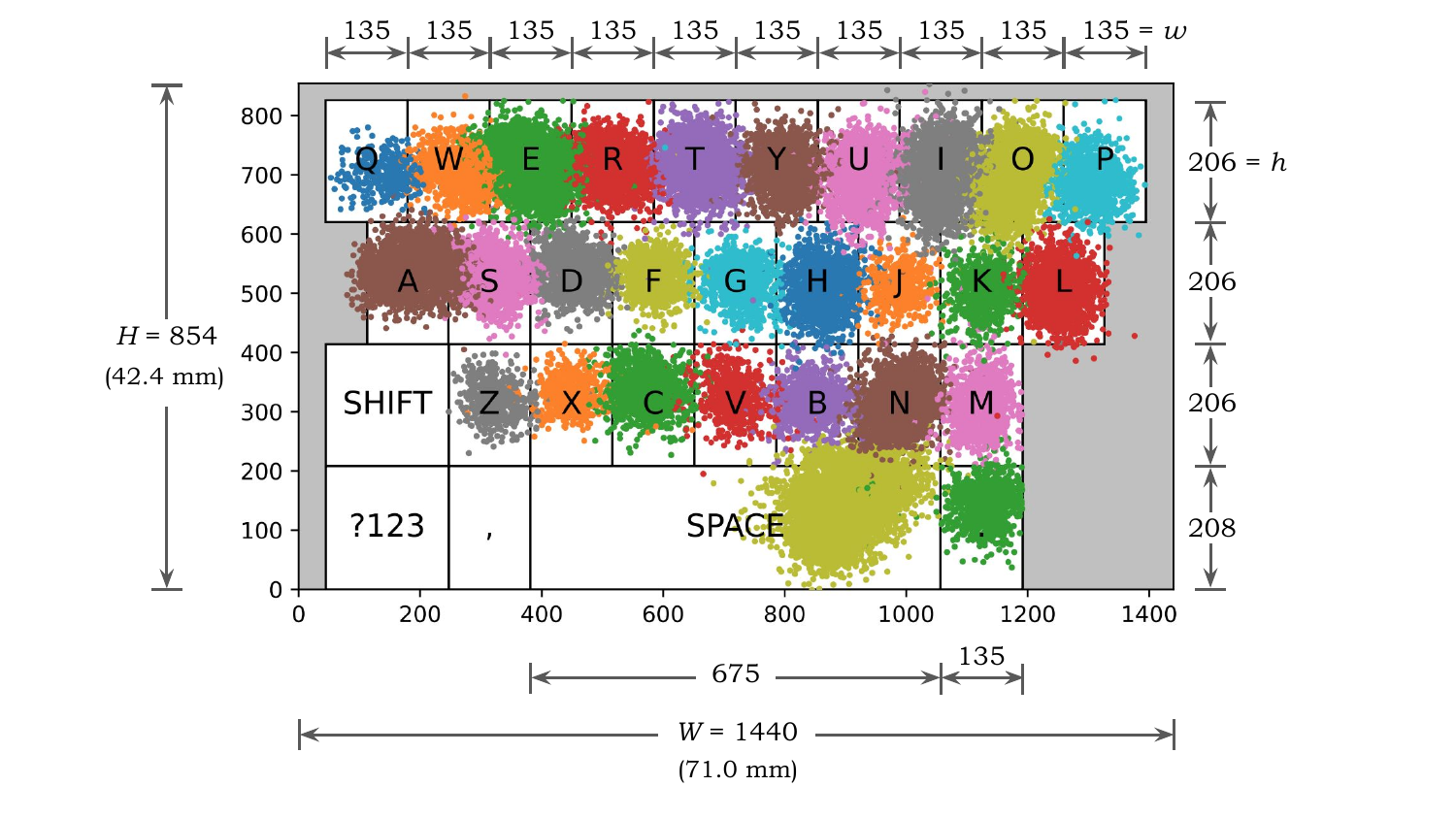}
  \caption{The distribution of touch centroids in the dataset from Study 1 pooled across all 24 participants. It is plotted on the QWERTY keyboard layout used in this study, which is the same layout as in the other two studies. The keyboard width ($W$), the keyboard height ($H$), the most common key width ($w$), and the most common key height ($h$) in pixels of the keyboard layout are also annotated in this figure. \textit{Resolution scale}: 1 pixel $\approx$ 0.05 mm.  } \label{fig:centroid_distribution}
  \Description{The QWERTY keyboard layout with clusters of dots representing touch centroids. Each cluster has a different color showing touch centroids of a specific reference key. The keyboard width ($W$ = 1,440 pixels), the keyboard height ($H$ = 854 pixels), the most common key width ($w$ = 135 pixels), and the most common key height ($h$ = 206 pixels) of the keyboard layout are also annotated in this figure.}
\end{figure}

The devices used in our studies were Pixel 6 Pro with the keyboard layout shown in Fig.~\ref{fig:centroid_distribution}.
The figure also has the keyboard layout dimensions annotated, including the keyboard width ($W$ = 1,440 pixels), the keyboard height ($H$ = 854 pixels), the most common key width ($w$ = 135 pixels), and the most common key height ($h$ = 206 pixels), which will be used next.
To assess the contribution of touch heatmaps to key decoding, we trained logistic regression-based spatial models that take a centroid and/or a heatmap as input and predict the probabilities of the candidate keys. In this paper, we focus on 28 candidate keys ($K = 28$) including the 26 English letters, SPACE, and PERIOD. Any accuracy difference observed between the models should reflect the contribution added by the heatmaps.

\subsection{Feature Engineering}
Our studies concern two types of features -- the touch centroid and the touch heatmap.
For \textbf{the touch centroid (\cf)} at the position $(x, y)$, we represent it with $28 \times 2$ numbers which are $[\Delta x_1, \Delta y_1, \Delta x_2,$ $\Delta y_2, \ldots, \Delta x_{28}, \Delta y_{28}]$ where
\begin{equation}
    \Delta x_k = \frac{x-x_k}{w} \mbox{\quad and \quad} \Delta y_k = \frac{y-y_k}{h}.
\end{equation}
$\Delta x_k$ and $\Delta y_k$ are the normalized signed distances from the touch centroid $(x, y)$ to the center of the $k^{th}$ key $(x_k, y_k)$ along the $x$ and the $y$ axes, respectively. 
As displayed in Fig.~\ref{fig:centroid_distribution}, $w$ is the most common key width in the keyboard layout (135 pixels), while $h$ is the most common key height (206 pixels).
Finally, we perform min-max normalization on all the $\Delta x_k$ and $\Delta y_k$ so the feature values stay in the range of $[-1, 1]$.

\begin{figure}[t]
  \centering
  \includegraphics[trim={2cm 3.25cm 9cm 7.5cm},clip,width=0.98\linewidth]{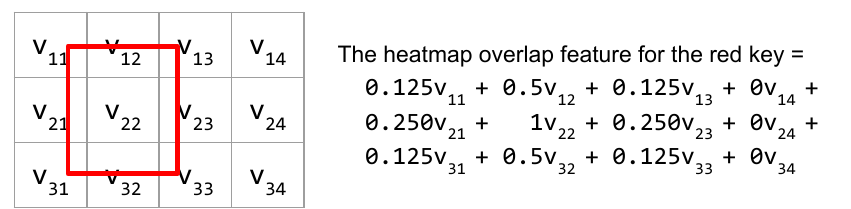}
  \caption{An illustration of how to compute the heatmap overlap feature for a given key. The key boundary is represented by the red box, while the heatmap sensor array is represented by the grid the red box is on. $v_{ij}$ represents the intensity value of the heatmap cell at row $i$ and column $j$. 
  The heatmap in this figure is made smaller ($3\times4$) than its
  actual size for clarity of illustration.} \label{fig:hov_feature}
  \Description{A three-by-four grid with a red box overlaying on it. The red box covers some parts of the first three cells of each row. On the right side, there is an equation showing how to compute the heatmap overlap feature for the key.}
\end{figure}

For touch heatmaps, each frame is a single-channel image of size $39 \times 18$ from the PCT sensors. However, we use only the lower part of the image (i.e., the last 16 rows) that covers the keyboard area, as displayed in Fig.~\ref{fig:heatmap_example} for example, for efficient computing. 
After exploring a number of alternatives, we chose the following two heatmap feature representations in our empirical experiments presented in this paper: \textbf{the flattened heatmap (\hf)} and \textbf{the heatmap overlap vector (\hov)}. 
The flattened heatmap (\hf) is obtained by flattening the two-dimensional ($16 \times 18$) heatmap intensity array into a vector of size 288 in row-major order.
In contrast, the heatmap overlap vector (\hov) represents the heatmap as a vector $f$ of size 28 corresponding to the 28 candidate keys. 
Each value is the weighted sum of the intensities of the heatmap cells that overlap with the area of the corresponding key. Mathematically,
\begin{equation}
    f_k = \sum_{i=1}^{16}\sum_{j=1}^{18} \frac{O(k, i, j)}{A_k}v_{ij}.
\end{equation}
where $f_k$ is the value in the heatmap overlap vector that corresponds to the $k^{th}$ candidate key (which has the area $A_k$ in the keyboard layout), $v_{ij}$ is the intensity of the heatmap cell at row $i$ and column $j$, and $O(k, i, j)$ is the overlapping area between the $k^{th}$ candidate key and the heatmap cell at row $i$ and column $j$. An illustration of the calculation is shown in Fig.~\ref{fig:hov_feature}. As with the centroid, we perform min-max normalization on both the flattened heatmap and the heatmap overlap vector so the feature values stay in the range of $[-1, 1]$. 

\subsection{Model and Training} \label{subsec:logreg}

\begin{figure}[t]
  \centering
  \includegraphics[width=0.99\linewidth]{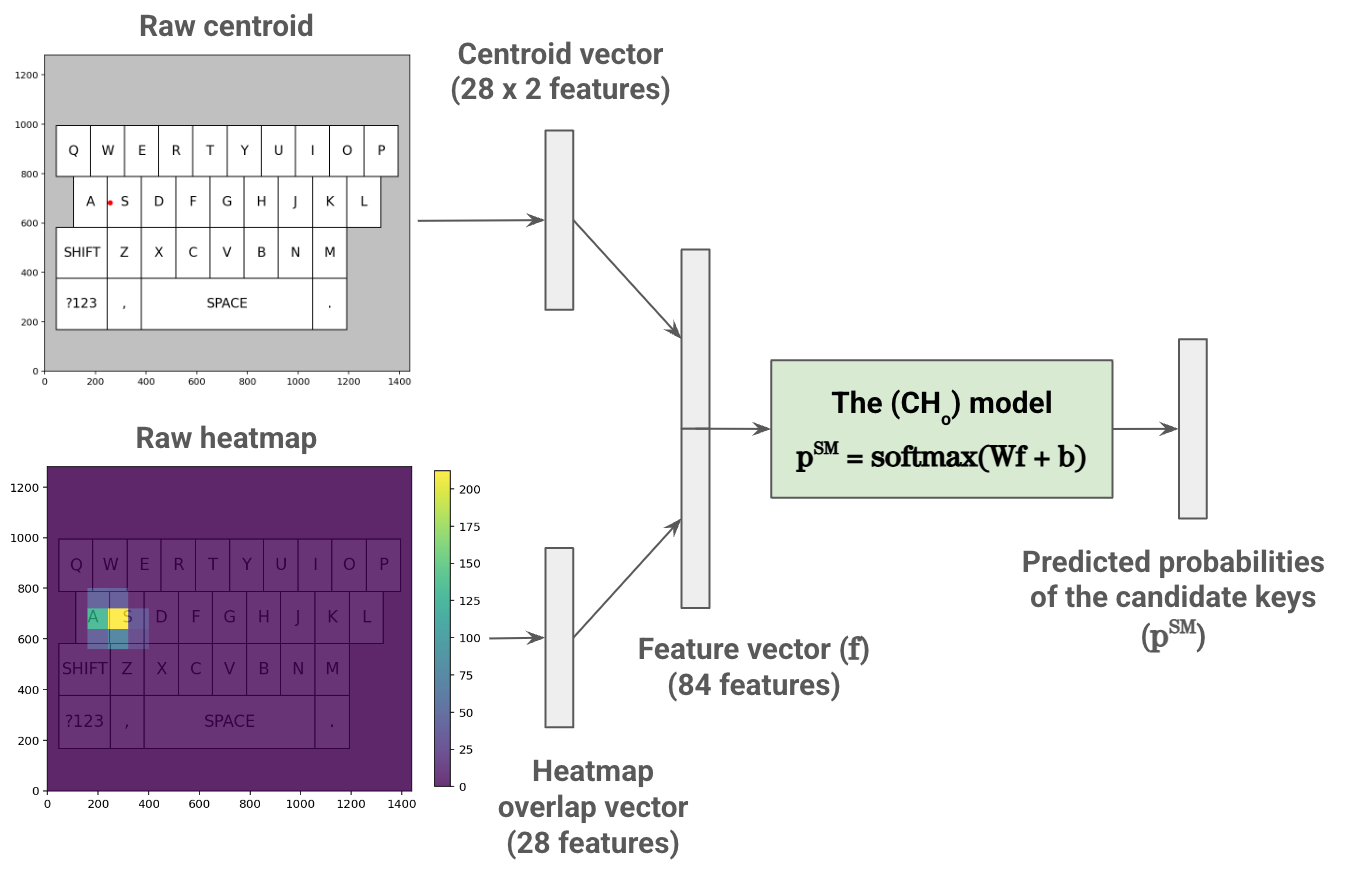}
  \caption{The (\chov) logistic regression model takes the centroid and the heatmap overlap vector as input and predicts probabilities of the candidate keys. Note that $W$ and $b$ are trained parameters of the model.
  } \label{fig:cho_diagram}
  \Description{A diagram shows how the logistic regression model that takes the centroid and the heatmap overlap vector as input works. The top-left part shows the derivation of the centroid input vector from the raw centroid, while the bottom-left part shows the derivation of the heatmap overlap vector from the raw heatmap. Then the centroid input vector and the heatmap overlap vector are concatenated and fed to the multi-class logistic regression model to predict the probabilities of the candiate keys.}
\end{figure}

We use multi-class logistic regression models as spatial models for predicting probabilities ($p^{SM}$) of the $K$ candidate keys. 
Mathematically, 
\begin{equation}
    p^{SM} = \mbox{softmax}(Wf + b)
\end{equation}
where $p^{SM} \in [0,1]^K$, $f \in \mathbb{R}^d$ is the feature vector of size $d$, and $W \in \mathbb{R}^{K \times d}$ and $b \in \mathbb{R}^K$ are model parameters. 
If the model uses both the centroid and the heatmap as input, we concatenate the two feature vectors into one before feeding it to the logistic regression model. 
Figure~\ref{fig:cho_diagram} illustrates this process for the model that takes the centroid and the heatmap overlap vector as input.

With the scikit-learn library 1.0.2 \cite{scikit-learn}, we train the models using the categorical cross-entropy loss $L_{CE}$ and the $L_2$ regularization loss: 
\begin{align}
    L_{CE} &= -\frac{1}{N}\sum_{i=1}^N\sum_{k=1}^Ky_{i,k}\mbox{log}(p_{i,k}^{SM})\\
    L_2 &= \frac{1}{2}\Vert W \Vert^2_F = \frac{1}{2}\sum_{k=1}^K\sum_{j=1}^d W^2_{k,j}
\end{align}
where $N$ is the number of training examples, $p_{i,k}^{SM}$ is the predicted probability of the candidate key $k$ for the $i^{th}$ example, and $y_{i,k}$ equals 1 if and only if the label of the $i^{th}$ example is the key $k$ (otherwise, 0).
Together, the final loss is
\begin{equation}
    L = L_{CE} + \frac{1}{C}L_2
\end{equation}
where $C$ is a hyperparameter called the inverse of regularization strength.
The LBFGS solver \cite{zhu1997algorithm} was used to optimize the models, while we selected $C$ from \{0.5, 1, 1.5, 2.0\} that led to the best validation accuracy. Training stopped upon parameter convergence or after 1000 iterations.
More details about the data splits and the feature sets will be explained in Section~\ref{sec:modelingstudy}.

\subsection{Additional Techniques} \label{subsec:additionaltechniques}

In addition to the study of the spatial models themselves, we tested their interaction effect with a set of three additional techniques used during decoding for further optimizing the typing experience.

\subsubsection{Language Model (LM)} \label{subsec:lm}
Combining the spatial model with a LM usually leads to better key decoding accuracy \cite{goel2013contexttype,goodman2002language,sivek2022spatial}. 
Here, we used a finite-state transducer (FST) LM, described in \cite{ouyang2017mobile}, which has been employed in several mobile text entry situations (e.g., \cite{hellsten2017transliterated,li2023c,sivek2022spatial}). This model predicts the probability of the next character given a previous context of at most five words.
However, this LM has two limitations. 
First, its training data did not include the PERIOD key. 
Second, its accuracy for the first character of each word could be low since the prompt set we constructed, to be explained in Section~\ref{subsec:touch-data-collection}, contains phrases with uncommon words to balance the character unigram distribution.
Hence, we decided not to rely on this LM for these two cases.
In other words, the logic we used for combining spatial model scores with language model scores is as follows. 

\begin{itemize}
    \item \textbf{If} the spatial model predicts PERIOD \textbf{or} the key to predict is the leading character of a word:
    \begin{itemize}
        \item \textbf{Answer} $\mbox{argmax}_k\{p^{SM}_k\}$
    \end{itemize}
    \item \textbf{Else}:
    \begin{itemize}
        \item \textbf{Answer} $\mbox{argmax}_k\{p^{SM}_k \times p^{LM}_k\}$
    \end{itemize}
\end{itemize}
where $p^{SM}_k$ and $p^{LM}_k$ are the spatial score and the LM score of the candidate key $k$, respectively. In our experiments, the $p^{SM}$ was the key probability predicted by the logistic regression model in Section~\ref{subsec:logreg}, where the $p^{LM}$ came from the FST model explained above.

\subsubsection{Skipping Unambiguous Cases (SUC)} 
Even though the user accurately taps on the target key, the final predicted key could be another key if the language model signal is so strong that it dominates the spatial signal. This might cause surprises and adversely affect the user experience.
To address this, we avoid using the spatial model to decode taps of which the touch centroid stays close to the center of a candidate key.
In such cases, we consider the taps unambiguous and predict the closest candidate keys directly instead.
Formally speaking, we consider a tap with the touch centroid $(x, y)$ unambiguous if there is a $k^{th}$ candidate key, with the key center $(x_k, y_k)$, such that $|x - x_k| < 0.25w_k$ and $|y - y_k| < 0.25h_k$ where $w_k$ and $h_k$ are the width and the height of the key $k$, respectively.
Note that this is similar to the concept of \textit{anchoring} in \cite{gunawardana2010usability}. 

\subsubsection{Neighbor Key Filtering} 
In order to increase the decoding accuracy and reduce surprises during decoding, we can limit the set of candidate keys to only the keys of which the touch centroid is not farther away from the key center than their immediate neighbor keys or in the equivalent distance. For example, the set of filtered candidate keys for the touch point in Fig.~\ref{fig:heatmap_example} is only \{s, d, f, z, x, c, SPACE\}. In this case, the filtering technique can reduce the number of candidate answers from 28 to only 7, mitigating the situation where the a strong signal from the language model sharply contradicts the spatial location of the touch point.

\subsection{Baselines} \label{sec:baselines}

For evaluation, we compare the logistic regression models to two baselines for key decoding.
\begin{itemize}
    \item \textit{On-key}: Predict the key bounding box which the touch centroid falls into (if any). If the touch centroid does not fall into any key bounding boxes, however, predict the closest key. "Closest" here is determined by the Euclidean distance between the touch centroid and the key center. This method returns only a categorical output without calculating the key probabilities.
    \item \textit{Distance}: Always predict the candidate key with the minimum normalized distance between the key center and the touch centroid, no matter which key the centroid is on. 
    We calculate the normalized distance to the $k^{th}$ key as follows.
    \begin{equation}
        d_k = \sqrt{\left(\frac{x-x_k}{W}\right)^2 + \left(\frac{y-y_k}{H}\right)^2}
    \end{equation}
    where $(x,y)$ is the touch centroid, $(x_k, y_k)$ is the center of the $k^{th}$ key, $W$ is the keyboard width (1,440 pixels), and $H$ is the keyboard height (854 pixels).
    Moreover, we can derive the predicted probabilities of the candidate keys (i.e., the $p^{SM}_k$ in Section~\ref{subsec:additionaltechniques}) based on the distances.
    Specifically, each distance $d_k$ is used as an input to a 1D Gaussian distribution to compute a probability density function (pdf) score $s_k$. 
    \begin{equation}
        s_k = \frac{1}{\sqrt{2\pi\sigma^2}} \mbox{exp}\left(\frac{-d_k^2}{2\sigma^2}\right)
    \end{equation}
    Then $s_k$ is normalized by the sum of the scores from all the candidate keys to be the key probability $p^{SM}_k$.
    \begin{equation}
        p^{SM}_k = \frac{s_k}{\sum_j s_j}
    \end{equation}
    The value of $\sigma$ is obtained empirically. We used $\sigma=0.03$ which optimized the categorical cross entropy loss on all validation splits. 
\end{itemize}

Because the Distance baseline (and the On-key baseline sometimes) simply use the distance from key centers to make prediction, we require a special way to treat the SPACE key which has a wider key width than other keys.
In our keyboard layout, the width of the SPACE key is 675 pixels, whereas the common key width is 135 pixels (as displayed in 
Fig.~\ref{fig:centroid_distribution}). 
To treat the SPACE key similarly to other keys, we consider the distance from the SPACE key along the x-axis to be zero if the x-coordinate of the touch centroid stays in the range of $[\frac{135}{2}, 675-\frac{135}{2}]$-th pixel from the left edge of the SPACE key.
In other words, we start measuring the $\Delta x$ from the SPACE key from the inner-left and the inner-right boundaries instead of from the key center.

\section{Study 1 - Data collection and model training} \label{sec:modelingstudy}
In the first study, our goal was to collect touch heatmaps (along with
touch centroids for comparison) when
users tap-typed phrase with known ground truths. This novel dataset was then
used to train and compare machine-learning models that predict keys based on input heatmaps and/or
centroids.
These models form the basis of the offline and online evaluation in the Studies 2 and 3
to be described in following sections.

\subsection{Touch Data Collection} \label{subsec:touch-data-collection}
We set up a copy-typing task to collect touch data from 24 participants. 

\paragraph{Participant inclusion criteria.}
This study recruited participants from the public who are familiar with typing on a mobile phone. Additional inclusion criteria are:
\textit{(i)} using English as the primary language for mobile typing and
\textit{(ii)} having no significant motor or visual impairments that may affect the performance on the typing task.

\paragraph{Task details.} The task was divided into three blocks each of which contains 30 prompts. A prompt is a reference / target sentence or a phrase of length 3-6 words that a user is asked to type.
The participants were asked to copy-type the prompts using two-thumb typing while holding the smartphone with both hands in a seated posture. 
In the first two blocks, we asked them to use their usual fast typing speed while trying to maintain accuracy. 
In the last block, we asked them to type with an extremely fast speed without caring much about accuracy so we could obtain more challenging tap input.
Throughout the study, the participants may edit typing errors before submitting the text but this was not required unless the edit distance between the submitted text and the prompt was too high (i.e., greater than 60\% of the prompt length).
They might also take a break for two minutes between consecutive blocks.

\paragraph{Device and keyboard setup.}
For consistency and time-efficiency, we used two Pixel 6 Pro devices for data collection. Each participant used one of the phones provided by the experimenter. These phones were held in the default portrait orientation and configured to log 
touch heatmaps during the study.
We used an Android application for showing prompts and collecting submitted texts and touch data.
A custom build of Gboard was used in this app.
The median of the time between two consecutive heatmap frames logged was around 4.22 ms. This translates to a heatmap capture rate of $\sim$237 frames per second.
To collect as many touch points as possible, we disabled intelligent keyboard features such as next-word prediction,  double-space period, and grammar check. 
We also disabled the per-key visual and haptic feedback in the keyboard in order not to distract the participants.
However, to allow fast typing, auto correction was enabled, and the suggestion bar was shown so the user could expect the result after auto-correction. Nonetheless, we strongly discouraged the user from tapping the suggestions; otherwise, we would get fewer touch points than desired.

\begin{figure}[t]
  \centering
  \includegraphics[width=0.99\linewidth]{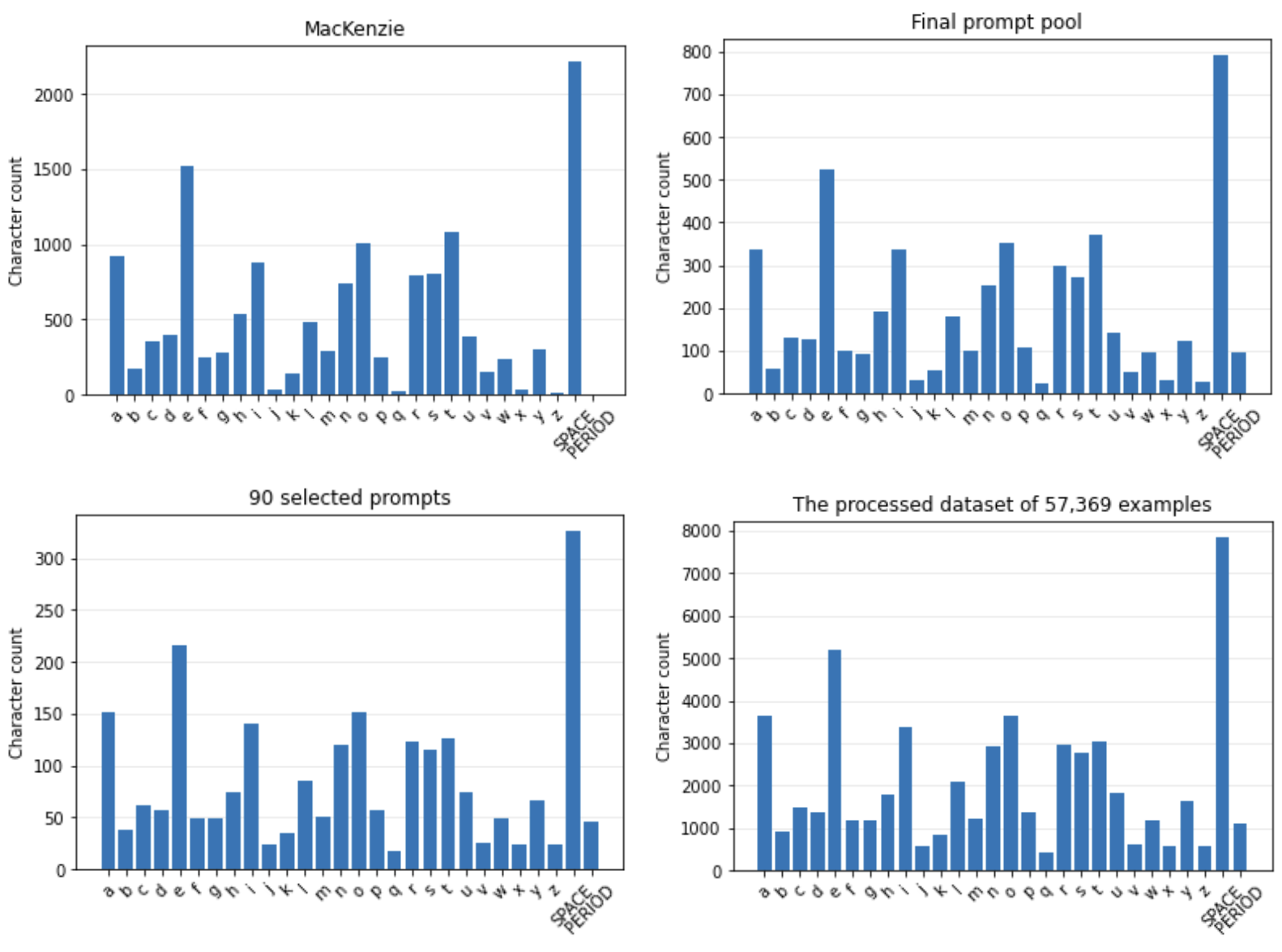}
  \caption{Character distributions (from A to Z and SPACE, PERIOD on the x-axis) in the MacKenzie and Soukoreff corpus \cite{mackenzie2003phrase} (top left), the final prompt pool for greedy selection (top right), the 90 selected prompts used for data collection (bottom left), and the processed main dataset used for training and evaluation (bottom right).} \label{fig:unigram_distribution}
  \Description{Four bar plots showing character distributions (character counts) of the prompt sets in the caption. The x-axes have the lower English alphabet (26 characters) plus SPACE and PERIOD.}
\end{figure}

\paragraph{Prompt set construction.}
We created a pool of prompts from the MacKenzie and Soukoreff phrase set \cite{mackenzie2003phrase}, which is widely used in the text-entry literature primarily for their ease to memorize in copy typing experiments (e.g., \cite{hincapie2014consumed,hoggan2008investigating,yu2017tap}).
However, the character distribution in this phrase set is extremely imbalanced, as shown in Fig.~\ref{fig:unigram_distribution} (top left). 
Sampling only from this phrase set would yield few touch points for rare characters (such as j, q, x, and z), insufficient for training our machine learning models.
To alleviate this problem, we added to the pool 30 new phrases which contain more rare characters than normal phrases.
In addition, to allow fast typing, we excluded the sentences with numbers, punctuation, and apostrophes, as they require switching to the secondary keyboard layout, long-pressing keys, or utilizing word suggestions which we discourage. Yet periods (\texttt{.}), which can be typed with the PERIOD key,
were allowed and appended to prompts that are sentences. Also, we tried to use only prompts that are not too complex by specifying the maximum numbers of words and rare words in each prompt.
Specifically, a word is considered rare if it is not in the list of 50k most common English words\footnote{according to the list most common English words, in order of frequency, in \\\url{https://github.com/arstgit/high-frequency-vocabulary}.}.
For general MacKenzie and Soukoreff phrases, we consider a phrase simple if it has at most 6 words and all of the words are not rare.
For each of the 30 phrases we manually added, we ensured that the total number of words + the number of rare words in the phrase is at most 6. Hence, if a phrase contains rare words, the maximum number of words must be more limited.
After filtering out overly complex phrases, we obtained the final prompt pool of which the character distribution is shown in Fig.~\ref{fig:unigram_distribution} (top right).

Finally, we constructed a set of selected prompts using greedy selection.
In other words, we started from an empty set. For each iteration, we selected a prompt from the pool that, if added to the set, would maximize the character-level entropy of the whole set. 
Once 90 prompts was reached (see the character distribution in Fig.~\ref{fig:unigram_distribution} (bottom left)), we randomly divided them into three subsets which were then used for the three task blocks of each participant. The three subsets were rotated using the Latin square design across all the 24 participants. 
Based on the prompts, each participant should contribute 2,379 taps. However, the actual number of examples we can use could be 
different from this
since we may gain additional taps from deleted touch points but also lose some due to unsuccessful alignment (see Section~\ref{subsec:alignment}).

\paragraph{Data collected.}
The touch centroid (x and y coordinates on the keyboard),
the touch heatmap (of size $39 \times 18$), and the timestamp of every raw touch event were collected along with the committed string.
As one tap may generate a sequence of touch heatmap events, depending on how long the finger contacts the screen, we used the data from the first frame of the sequence for training and evaluation. 
Practically, the keyboard needs to respond visually to the user taps by highlighting the pressed keys on finger-down. Therefore, using the first frame data allows the keyboard to provide the earliest possible response to a given user tap.

\subsection{Touch Point - Key Alignment} \label{subsec:alignment}
After obtaining the touch points, we aligned them to the characters in the prompt to create input-output pairs for training and evaluation. We could divide the alignments into two cases.

\paragraph{Committed touch points.} Here, we aligned touch points in the committed string to the reference prompt. 
Since the committed string may contain errors, we used the Needleman–Wunsch algorithm that supports insertion, omission, substitution, and transposition errors for alignment \cite{needleman1970general}. For words in the committed string that are the results of auto correction, we aligned the prompt to the corrected form first and then aligned the corrected form to the original form of which we have touch data. Using this chain of alignment, we could get pairs of touch data and reference keys.   

\paragraph{Deleted touch points.} Deleted touch points are very important in this work because they are the cases for which the 
keyboard we used during data collection failed to generate decoding results expected by the user, showing room for improvement. To infer the intended keys of the deleted touch points, we replayed the typing sequence (including backspaces) step by step and did the alignment, using the Needleman–Wunsch algorithm, every time before a touch point was deleted. The reference text for alignment was a prefix of the prompt (with 1 character longer than the current text so far to allow an omission error). We only kept alignments of those not being aligned during committed touch point alignment.  

Aligning deleted touch points is sometimes imperfect since we cannot know exactly why the user deleted the touch points. This results in ambiguous cases where we have more than one 
alignment hypothesis.
For example, given the prompt \texttt{"Breathing is difficult"}, a user typed \texttt{"Be"} and then deleted \texttt{e}. 
It could be argued that the user intentionally typed \texttt{e} and forgot \texttt{r}, so this is a spelling error (omission), which is not the focus of this study. Otherwise, the user might in fact want to type \texttt{r} but slightly missed it to the left and the keyboard interpreted the user touch as \texttt{e}, a neighboring key of \texttt{r}, instead. This is, in contrast, a spatial error we are interested in. 
Another example is, for the prompted word \texttt{"missed"}, the user typed \texttt{"missd"} and then deleted \texttt{d}. The deleted \texttt{d} could be aligned to both \texttt{e} and \texttt{d}.
As we do not know which alignment is indeed the case, we detected examples of this category, i.e., spelling (omission) error vs spatial error, and excluded them from our dataset.

Lastly, the key from the alignment might be too far from the touch centroid on the QWERTY keyboard. 
For example, given the prompt \texttt{"Buffer zones near Iraq"}, a user may type and submit \texttt{"Buffer zones near Iran"} because they did not pay attention to the prompt carefully. The algorithm aligns the user's tap of \texttt{n} to the reference character \texttt{q}. However, we know that this is not a spatial error we are interested in since the touch centroid (near \texttt{n}) and the key \texttt{q} are too far on the keyboard. So, we excluded this case from our dataset.
Generally speaking, we kept only examples where the touch centroid stays no farther than the immediate closest keys of the reference key or in the equivalent distance.

\subsection{Statistics of Touch Data} \label{subsec:datastat}

\begin{table*}[t]
  \caption{A summary of the datasets used in our modeling, offline evaluation, and online evaluation studies. M\&S denotes prompts from the MacKenzie and Soukoreff corpus \cite{mackenzie2003phrase} with periods appended to prompts that are sentences.}
  \label{tab:datasets}
  \begin{tabular}{L{2.5cm} L{3.75cm}  R{2.6cm} R{2.1cm} R{2.3cm} R{2.3cm}}
    \toprule
    Dataset & Prompt styles & No. of trials & No. of participants & No. of tap examples & Used for evaluation in\\
    \midrule
    Main & M\&S + Phrases with more rare characters & \makecell[r]{30 prompts/block \\ $\times$ 3 blocks} & 24 &  57,369 & Section~\ref{subsec:id}\\ 
    OfflineTest-Subset & A subset of prompts used in Main & \makecell[r]{15 prompts/block \\ $\times$ 1 block} & 10 & 4,096 & Section~\ref{subsec:offlinetest-subset}\\ 
    OfflineTest-Unseen & M\&S (not overlapping with Main) & \makecell[r]{15 prompts/block \\ $\times$ 1 block} & 10 & 4,571 & Section~\ref{subsec:offlinetest-unseen}\\ 
    OnlineTest & M\&S (not overlapping with Main) + OOV strings & \makecell[r]{30 prompts/block \\ $\times$ 4 blocks} & 16 & 43,735 & Section~\ref{subsec:online-results}\\ 
    \bottomrule
\end{tabular}
\end{table*}

After processing the collected data as explained in Section~\ref{subsec:alignment}, we obtained 57,369 examples. 56,492 of them are committed touch points, while the rest (877 examples) are deleted touch points.
On average, a participant contributed 2,390.38 examples (standard deviation: 23.33 examples). The smallest and the largest numbers of examples a participant contributed were 2,333 and 2,440 examples, respectively.
The distribution of the reference key labels, aggregated from all participants, is displayed in Fig.~\ref{fig:unigram_distribution} (bottom right), while the distribution of the touch centroids on the keyboard layout is displayed in Fig.~\ref{fig:centroid_distribution}.
Overall, we call this dataset ``Main'' in Table~\ref{tab:datasets}.

\subsection{Model Training and Evaluation} \label{subsec:id}
\subsubsection{Models}
To investigate the contribution of the touch heatmap, we compared the logistic regression-based spatial models of five input feature sets including
\begin{itemize}
    \item \textbf{(\cf) using only the touch centroid as input}
    \item \textbf{(\hf) using only the flattened heatmap as input}
    \item \textbf{(\chf) using both the touch centroid and the flattened heatmap as input}
    \item \textbf{(\hov) using only the heatmap overlap vector as input} 
    \item \textbf{(\chov) using both the touch centroid and the heatmap overlap vector as input.}
\end{itemize}
If the model uses both the centroid and the heatmap features, we concatenate their feature vectors to be the input of the logistic regression model. 
For example, the (\chov) model has 56 (from \cf) + 28 (from \hov) = 84 features in total.
Additionally, we compared these models to the two baselines, On-key and Distance, which require no training.

\subsubsection{Setup}
We performed the leave-one-out cross validation on the 24 participants in our main dataset.
Specifically, for each fold, we used examples from one participant to be the test set. For the remaining 23 participants, we used the examples from 20 participants to be the training set and the other 3 participants to be the validation set.
This validation set was used to choose the best hyperparameter for the logistic regression model, i.e., choosing the best inverse of regularization strength from the search space of \{0.5, 1.0, 1.5, 2.0\}.
The cross-entropy losses for training and validation are reported in Appendix~\ref{app:losses}.

We tested all the models under two settings -- without and with the additional techniques in Section~\ref{subsec:additionaltechniques}.
The main metric used was the character error rate averaged from the 24 test participants.
We define character error rate (CER) as the percentage of keypresses that are wrongly decoded by the model according to the aligned data. Lower CER indicates fewer misinterpretations of user taps, leading to fewer typing errors and tending to improved speed and user experience.
Significance tests between two methods were conducted using paired sample tests (i.e., considering the difference for each participant). 
Specifically, for the differences that were normally distributed according to the normality test \cite{d1973tests}, we used paired t-test for the significance test; otherwise, we used the Wilcoxon signed-rank test \cite{wilcoxon}, which better supports the non-normality distribution. 
The significance level $\alpha$ used was 0.05 for all the tests.

\begin{table}[t]
  \caption{Character error rates (CERs) on the main dataset when using the spatial model (SM) only and when the additional techniques were incorporated.
  \cf, \hf, and \hov denote the touch centroid, the flattened heatmap, and the heatmap overlap vector input features, respectively, while \chf is \cf and \hf concatenated and \chov is \cf and \hov concatenated.
  Bold numbers indicate the least error rate achieved in each setting. 
  Underlined numbers are for the winner method and those that are not significantly different from the winner in each setting. 
  }
  \label{tab:id}
  \begin{tabular}{L{1.4cm} R{1.55cm}  R{1.75cm} R{2.3cm}}
    \toprule
    \multicolumn{2}{l}{Leave-one-out CV} & \multicolumn{2}{c}{CER $\downarrow \pm$ standard deviation (\%)}\\
    \multicolumn{2}{l}{(Section~\ref{subsec:id})} & SM only & + Add. techniques\\
    \midrule
    \multicolumn{2}{l}{\textit{Baselines}} && \\
    \multicolumn{2}{l}{\quad On-key} & 5.67 $\pm$ 2.13 & - \\
    \multicolumn{2}{l}{\quad Distance} & 5.59 $\pm$ 2.16 & 2.50 $\pm$ 1.01 \\
    \midrule
    \multicolumn{2}{l}{\textit{Logistic Regression}} && \\
    \quad (\cf) & 56 features & 4.22 $\pm$ 1.79 & 2.87 $\pm$ 0.95\\
    \quad (\hf) & 288 features & \textbf{\underline{3.31 $\pm$ 1.29}} & \textbf{\underline{2.02 $\pm$ 0.88}}\\
    \quad (\chf) & 344 features & \underline{3.32 $\pm$ 1.26} & \underline{2.04 $\pm$ 0.88}\\
    \quad (\hov) & 28 features & 3.77 $\pm$ 1.61 & 2.27 $\pm$ 0.97\\
    \quad (\chov) & 84 features & 3.64 $\pm$ 1.52 & 2.23 $\pm$ 0.94 \\
    \bottomrule
\end{tabular}
\end{table}

\subsubsection{Results of the Spatial-Only Models}
Table~\ref{tab:id} reports the character error rates (CERs) averaged from the 24 participants. 
Looking at the ``SM only'' column where we consider the raw prediction from the spatial models, we notice that, first, the On-key and the Distance baselines performed worse than the other methods that leverage training data.
Next, the (\chf) model further reduced the character error rate from the (\cf) model by 0.90\% absolute or 21.4\% relative, while the (\chov) model could reduce it by 0.58\% absolute or 13.7\% relative.
These highlight that \textbf{the information from touch heatmap can enhance the performance of the spatial model for key decoding}.
Moreover, since (\hf) was better than (\hov) and (\chf) was better than (\chov), we can conclude that using the flattened heatmap yielded lower error rates than the heatmap overlap vector for in-distribution data.   
Last but not least, for \hf, using the heatmap alone was slightly better than using it together with the centroid (paired t-test: $t(23)=0.3399, p=0.74$). 
Meanwhile, for \hov, using the heatmap alone  was significantly better than using it together with the centroid (Wilcoxon signed-rank test: $W=30, p=2.39\times10^{-4}$).
\footnote{The effect sizes of the results of all the three studies are reported in Appendix~\ref{app:effectsizes}.}

\begin{figure}[t]
  \centering
  \includegraphics[trim={1cm 0.5cm 1cm 1cm},clip,width=0.98\linewidth]{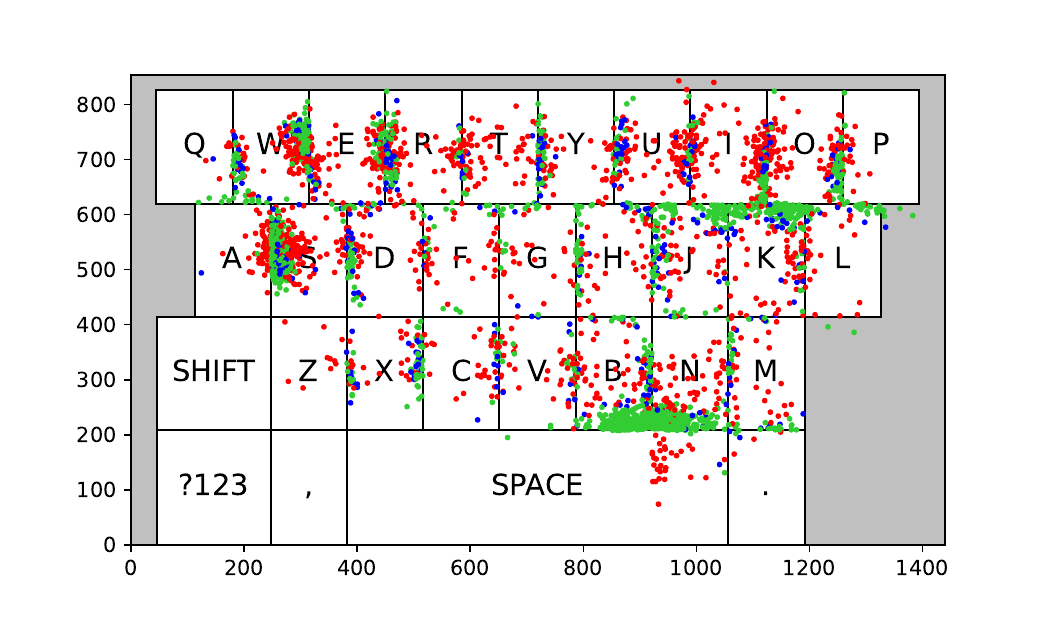}
  \caption{Touch centroids of misclassified examples of the (\chf) model and/or the Distance baseline when using the spatial models only. Red dots are touch centroids of examples misclassified by both the (\chf) model and the Distance baseline. Green dots are of examples classified correctly by the (\chf) model but misclassified by the Distance baseline. Blue dots are of examples classified correctly by the Distance baseline but misclassified by method the (\chf) model.}  \label{fig:misclassified-top}
  \Description{A scatter plot of touch centroids of misclassified examples on the keyboard layout. The dots are dense around key boundaries.}
\end{figure}

\begin{figure}[t]
  \centering
  \includegraphics[trim={1cm 0.5cm 1cm 1cm},clip,width=0.98\linewidth]{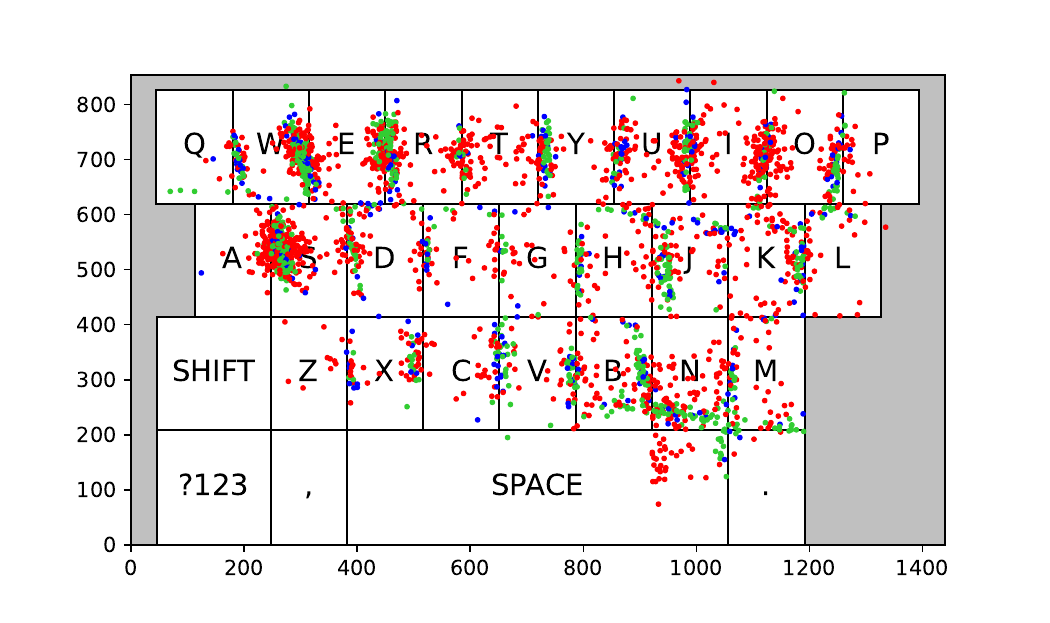}
  \caption{Touch centroids of misclassified examples of the (\chf) model and/or the (\cf) model when using the spatial models only. Red dots are touch centroids of examples misclassified by both the (\chf) model and the (\cf) model. Green dots are of examples classified correctly by the (\chf) model but misclassified by the (\cf) model. Blue dots are of examples classified correctly by the (\cf) model but misclassified by method the (\chf) model.}  \label{fig:misclassified-bottom}
  \Description{A scatter plot of touch centroids of misclassified examples on the keyboard layout. The dots are still dense around key boundaries, but they are relatively more sparse than the previous figure.}
\end{figure}

To better understand the strengths and the weaknesses of the spatial models, we plot touch centroids of misclassified examples of some models in Figs.~\ref{fig:misclassified-top}-\ref{fig:misclassified-bottom}. Specifically, Fig.~\ref{fig:misclassified-top} compares the (\chf) model to the Distance baseline. The red dots are the examples which both methods misclassified. The green dots were correctly classified only by the (\chf) model but not the Distance baseline, whereas the blue dots show the opposite cases. 
As expected, most of these misclassified cases stay near the boundaries of the keys.
Also, we can see many green dots above the SPACE key (inside the keys B and N) and under the keys I and O. These are cases where the touch centroids strayed away from the intended keys but the (\chf) model can rescue them.
Examples that are challenging for both methods are those aimed to type A but the centroid was on the key S as well as the misclassifications between two consecutive keys on the first row of the keyboard layout (especially W-E, E-R, U-I, and I-O).
In addition, we notice some red dots inside the SPACE key, but the intended key was actually the PERIOD key. 
We hypothesize that some participants might get used to tapping the SPACE key twice to add a period. However, this feature was disabled during data collection, resulting in the red dots. 
Fig.~\ref{fig:misclassified-bottom} compares the (\chf) model to the (\cf) model. We can notice that some green dots shown on Fig.~\ref{fig:misclassified-top} disappear as the (\cf) model can also predict them correctly. 
Still, the (\chf) model can outperform the (\cf) model because of its strengths to handle ambiguous cases between several key pairs such as E-R, O-P, H-J, K-L, and B-N, as well as the cluster N-M-SPACE-PERIOD.

\subsubsection{Results with Additional Techniques}
The column ``+ Add. techniques'' of Table~\ref{tab:id} shows the decoding results after the three additional techniques in Section~\ref{subsec:additionaltechniques} (LM integration, skipping unambiguous cases, and neighbor key filtering) were applied. This setting is more similar to keyboard decoding in production where the spatial model is not in use alone.
Note that we have no result for the On-key baseline under this setting since it cannot predict the key probabilities to be combined with the LM scores.

According to the result table, the three additional techniques helped reduce the character error rates for all the spatial models. 
The trends we found in the previous subsection mostly held but the absolute differences became smaller. 
For example, the (\chf) model was still significantly better than the (\cf) model with 0.85\% absolute CER difference compared to 0.90\% absolute when using SM only. 
Also, the (\hf) model was still better than the (\hov) model with the 0.25\% absolute CER gap, compared to 0.46\% absolute CER gap when using SM only.
The explanation for these smaller differences could be that the three additional techniques (including the LM, skipping unambiguous cases, and neighbor key filtering) could rescue many misclassifications of the weaker spatial model, but they were already correctly classified by the stronger spatial model, leading to the smaller performance gaps between the two.

Interestingly, however, while the best error rate belonged to the (\hf) model, the Distance baseline got the largest absolute CER reduction ($\approx3\%$) due to the additional techniques, while the logistic regression models got around 1.25\%-1.50\% absolute CER reduction.
We hypothesize that the multiplication between $p^{SM}_k$ and $p^{LM}_k$ in Section~\ref{subsec:lm} suits the Distance baseline more than the logistic regression models.
Actually, for the Distance baseline, this multiplication is in line with the noisy channel model for speech recognition \cite{jelinek1998statistical}, the core idea of which is applicable to tap decoding as well.
Meanwhile, we may need a co-optimization of the spatial model and the language model to best integrate both information sources via machine learning.
Besides that, other score combination strategies as well as model calibration \cite{guo2017calibration} are worth exploring for further reducing the error rates.
We consider these research directions future work as, in this paper, we focus mainly on the spatial model aspect in order to arrive at clear conclusions concerning touch heatmap effects.

\section{Study 2 - An Offline Model Evaluation} \label{sec:ood}

The goal of Study 2 was to evaluate the performance of the models
we developed in Study 1 (Section~\ref{sec:modelingstudy}) on their generalizability to \textbf{new users and unseen phrases}.

\subsection{Test Sets for Offline Evaluation} \label{subsec:ooddata}
We collected the test data from ten new participants whom we asked to complete three typing task blocks.
We used one of the blocks to collect test touch points for this study and reserved the other blocks for future studies. 
The block for this study consists of 30 prompts. Half of them were selected from the previous 90 prompts in a way that maximizes the character-level entropy of each prompt subset, while the other half were randomly selected from the phrases in the MacKenzie and Soukoreff corpus different from the previous 90 prompts.
This led to two sets of data, called \textit{OfflineTest-Subset} (for it contained a subset of phrases used during model training) and \textit{OfflineTest-Unseen} (for it consisted of phrases unseen during model training), respectively.
This data collection used the same device and keyboard setup as in Section~\ref{subsec:touch-data-collection} and the same touch point-key alignment algorithm as in Section~\ref{subsec:alignment}.
We summarized the details and the number of examples of these two offline test sets in Table~\ref{tab:datasets}.

\subsection{Models and Setup}
Due to the leave-one-out cross validation in Study 1 (Section~\ref{subsec:id}), we had 24 logistic regression models for each input feature set.
To calculate the average CERs for each test dataset, we applied each of the 24 models to the touch points from each test user and averaged the CERs from the 24 models to represent the model performance for this user. After that, we calculated the mean across ten users to be the model performance for the dataset.
As with Section~\ref{subsec:id}, we evaluated the models under both the spatial model only setting and the setting with the three additional techniques applied, and we used either the paired t-test or the Wilcoxon test for comparing two models depending on the normality of the data.

\begin{table}[t]
  \caption{Character error rates (CERs) on the OfflineTest-Subset dataset (Top) and the OfflineTest-Unseen dataset (Bottom) when using the spatial model (SM) only and when the additional techniques are incorporated.
  Bold numbers indicate the least error rate achieved in each setting. 
  Underlined numbers are for the winner method and those that are not significantly different from the winner in each setting. 
  }
  \label{tab:ood}
  \begin{tabular}{L{1.4cm} R{1.55cm}  R{1.75cm} R{2.3cm}}
    \toprule
    \multicolumn{2}{l}{OfflineTest-Subset} & \multicolumn{2}{c}{CER $\downarrow \pm$ standard deviation (\%)}\\
    \multicolumn{2}{l}{(Section~\ref{subsec:offlinetest-subset})} & SM only & + Add. techniques\\
    \midrule
    \multicolumn{2}{l}{\textit{Baselines}} && \\
    \multicolumn{2}{l}{\quad On-key} & \underline{6.96 $\pm$ 4.48} & - \\
    \multicolumn{2}{l}{\quad Distance} & \underline{6.84 $\pm$ 4.38} & \underline{3.85 $\pm$ 1.55} \\
    \midrule
    \multicolumn{2}{l}{\textit{Logistic Regression}} && \\
    \quad (\cf) & 56 features & \underline{6.68 $\pm$ 3.68} & 4.74 $\pm$ 1.69\\
    \quad (\hf) & 288 features & \underline{6.44 $\pm$ 3.80} & \textbf{\underline{3.78 $\pm$ 1.74}}\\
    \quad (\chf) & 344 features & \underline{6.38 $\pm$ 3.80} & 3.86 $\pm$ 1.79\\
    \quad (\hov) & 28 features & \underline{6.43 $\pm$ 3.56} & \underline{3.83 $\pm$ 1.48}\\
    \quad (\chov) & 84 features & \textbf{\underline{6.32 $\pm$ 3.74}} & \underline{3.78 $\pm$ 1.53} \\
    \bottomrule
\end{tabular}
\bigskip

\begin{tabular}{L{1.4cm} R{1.55cm}  R{1.75cm} R{2.3cm}}
    \toprule
    \multicolumn{2}{l}{OfflineTest-Unseen} & \multicolumn{2}{c}{CER $\downarrow \pm$ standard deviation (\%)}\\
    \multicolumn{2}{l}{(Section~\ref{subsec:offlinetest-unseen})} & SM only & + Add. techniques\\
    \midrule
    \multicolumn{2}{l}{\textit{Baselines}} && \\
    \multicolumn{2}{l}{\quad On-key} & 8.21 $\pm$ 4.92 & - \\
    \multicolumn{2}{l}{\quad Distance} & 8.25 $\pm$ 4.91 & \underline{4.11 $\pm$ 1.94} \\
    \midrule
    \multicolumn{2}{l}{\textit{Logistic Regression}} && \\
    \quad (\cf) & 56 features & 6.92 $\pm$ 4.27 & 4.52 $\pm$ 2.39\\
    \quad (\hf) & 288 features & 6.75 $\pm$ 4.20 & \underline{3.72 $\pm$ 2.13}\\
    \quad (\chf) & 344 features & 6.65 $\pm$ 4.13 & \underline{3.74 $\pm$ 2.16}\\
    \quad (\hov) & 28 features & \textbf{\underline{6.09 $\pm$ 3.76}} & \textbf{\underline{3.63 $\pm$ 2.17}}\\
    \quad (\chov) & 84 features & \underline{6.22 $\pm$ 3.76} & \underline{3.68 $\pm$ 2.28} \\
    \bottomrule
\end{tabular}

\end{table}

\subsection{Results from OfflineTest-Subset} \label{subsec:offlinetest-subset}
Table~\ref{tab:ood} (top) shows the results from the OfflineTest-Subset dataset.
Under the SM only setting, the (\chov) model performed the best, but the differences were not statistically significant from the other methods, possibly due to the small sample size (i.e., only ten participants).
With the additional techniques, the (\hf) model performed the best, comparably to the (\hov) model, the (\chov) model, and the Distance baseline.
However, we still see the clear benefit of using the heatmap features over the (\cf) model.

\subsection{Results from OfflineTest-Unseen} \label{subsec:offlinetest-unseen}
Table~\ref{tab:ood} (bottom) shows the results from the OfflineTest-Unseen set.
Considering the baseline results, we could say that OfflineTest-Unseen is more challenging to decode than OfflineTest-Subset. 
This is likely because
OfflineTest-Unseen does not contain the prompts with rare characters we added to balance the character distribution (see Section~\ref{subsec:touch-data-collection}). 
Therefore, for OfflineTest-Unseen, the participants could increase their speed more, leading to more challenging touch points.
However, under the SM only setting, the best method for OfflineTest-Unseen was the (\hov) model, followed by the (\chov) model. 
Moreover, the (\hov) model was significantly better than the (\hf) model, so we may say that \textbf{the heatmap overlap vector is more generalizable than the flattened heatmap for unseen prompts}.
After the additional techniques were applied, the performance gap between these two heatmap representations was narrowed down.
Still, the models that used heatmap features were significantly better than the (\cf) model and had a relatively large performance gap compared to the Distance baseline.

\section{Study 3 - Closing the Loop: An Online Interactive Model Evaluation} \label{sec:userstudy}

To assess the practical effectiveness of heatmap-based keyboard decoding
and how it compares with the machine-learning model based on centroids,
we deployed the (\hov) model and the (\cf) model to
smartphones and measured human users' live typing performances in Study 3.
This was, to our knowledge, the first-ever systematic investigation
of live typing behavior supported by touch heatmaps.\footnote{We publicly release the data of this study at \url{https://github.com/google-research-datasets/tap-typing-with-touch-sensing-images}.} 

\subsection{Setup for Online Evaluation}

To ensure the consistency between the training and inference of the models,
we used the same model of smartphone hardware (Pixel 6 Pro), keyboard size and layout (default),
and screen orientation (default portrait) as the modeling study. 
We implemented two models on Pixel 6 Pro to compare in the user study: the (\hov) model and the (\cf) model. 
Among the feature sets with \hf or \hov, we chose the (\hov) model because of its highest average accuracy when decoding unseen prompts, according to Table~\ref{tab:ood} (bottom), suggesting that it likely works well in a realistic setting.\footnote{We performed a simulation of this study for the other heatmap-based models and presented the results in Appendix~\ref{app:simulate-study3}.} 
Due to the small size of the logistic regression models, their on-device inference, supported by TensorFlow Lite, achieved negligible latencies on the Pixel 6 Pro devices.
When training the models, we randomly selected 20 participants and 4 participants from Study 1 to be the training and the validation sets, respectively. The training and validation losses are reported in Appendix~\ref{app:losses}.
Also, we applied the three additional techniques in Section~\ref{subsec:additionaltechniques} (i.e., the LM, skipping unambiguous cases, and neighbor key filtering) for smoother user experience, making the setting more similar to decoding in production.

\paragraph{Participant inclusion criteria}

A total of 16 participants (8 male, 7 female, 1 non-binary) were recruited for this user study. 
The inclusion criteria included: 1) regularly uses typing on a smartphone,
2) uses English as the primary language for mobile typing, and 3)
no vision or motor impairment that may affect the copy typing task in the user study.

\paragraph{Task details}

Each user-study participant was asked to perform four separate typing task blocks, each
consisting of 30 text prompts to be copy-typed. In two of the blocks,
the keyboard used the (\hov) model for key decoding, while in the remaining two blocks,
the keyboard used the (\cf) model. The order of the (\hov) and the (\cf) blocks
were counterbalanced across and blinded from all the participants. 
Here, the users were instructed to type as quickly and as accurately as possible (with their usual fast speed). 
As with Section~\ref{subsec:touch-data-collection}, they were allowed to correct typing errors before each submission, but this was not mandatory unless the difference between their submitted text and the original prompt exceeded 60\% of the prompt's length.
The same data collection app as Section~\ref{subsec:touch-data-collection} was used in this online user study.
Following each block, 
the users completed a short survey
form and answered questions regarding the intended and achieved typing speed and
their satisfaction with the typing experience in the just-completed block. 
They were also given a 1-2 minute break before proceeding to the next block. 
The total duration of the entire user-study
session varied between 30 and 50 minutes among the 16 participants.

\paragraph{Prompt set construction}

The 30 prompts presented as the text for copy-typing in each block were divided into
two subsets:

\begin{enumerate}
    \item \textbf{Unseen English phrases}: 20 phrases from the MacKenzie and Soukoreff sentences \cite{mackenzie2003phrase}, none of which was used to train or validate the two models. 
    The selection of these phrases tested generalizability of the  models for general English typing. 
    \item \textbf{Random strings}: 10 strings of random characters
    chosen from the 26 English letters, SPACE, and PERIOD. Each random string was exactly 8-character long and does not have SPACE at the beginning or at the end to avoid visual ambiguity. These strings are considered out-of-vocabulary (OOV) unlike the unseen English phrases.
\end{enumerate}

The English phrases and random strings were mixed and shuffled randomly in order
in each block.
The four blocks consisted of non-overlapping prompts. In addition, the prompts were randomized
in order among the participants. During the lab study, the auto-correct feature of 
the keyboard was disabled in order to prevent the typing of random strings from
triggering unwanted corrections.

\paragraph{Data collected}

The data collected included the touch heatmaps and centroids, their associated
timestamps, as well as the LM scores associated with each
tap event (if available), and the result of the online decoding. Based on these
data, we computed the character- and word-level error rates and the typing speed (in words per minute, WPM)\footnote{We follow the definition of WPM in \cite{arif2009wpm} where five characters are considered one word so that we can compare WPMs of English phrases and random (OOV) strings.} and
compared their values between the conditions.

\subsection{Results of Online Evaluation} \label{subsec:online-results}

\subsubsection{Typing Metrics}

\begin{table}[t]
  \caption{Typing metrics by prompt types of both model conditions averaged across 16 participants in the user study.}
  \label{tab:metrics}
  \setlength\tabcolsep{4.1pt}
  \begin{tabular}{L{0.9cm} R{1.5cm}  R{1.5cm} R{1.4cm} R{1.7cm}}
    \toprule
    Model & \multicolumn{2}{c}{WPM $\uparrow \pm$ std.} &  \multicolumn{2}{c}{WER $\downarrow \pm$ std. (\%)}\\
    & Phrase & OOV  & Phrase & OOV \\
    \midrule
    (\cf) & 34.32 $\pm$ 8.14 & 13.35 $\pm$ 1.89 & 5.28 $\pm$ 6.98 & 15.57 $\pm$ 12.68\\
    (\hov) & 36.13 $\pm$ 8.70 & 13.60 $\pm$ 2.23 & 3.60 $\pm$ 4.44 & 12.40 $\pm$ 11.79\\
    \bottomrule
\end{tabular}
\end{table}

As shown in Table~\ref{tab:metrics}, as expected, the participants exhibited lower average typing speed (WPM) and higher word error 
rate (WER) when typing the random (OOV) strings than typing the unseen English phrases.
In the meantime, \textbf{their observed typing behavior showed both higher speed (WPM) and
lower WER under the blocks supported by the (\hov) model than the ones supported
by the (\cf) model.}

Specifically, the relative gain of the typing speed (in WPM) due to the (\hov) model relative to the (\cf) model was $5.27\%$ 
and reached significance under the paired t-test ($t(15)=2.4790, p=0.026$, two-tailed, same below) for the English phrases.
For the OOV strings, this relative gain was smaller ($1.87\%$) and did not reach statistical significance ($t(15)=0.7099, p=0.489$).
Similarly, in terms of typing accuracy, the word error rate (WER) showed
a $31.82\%$ reduction ($3.60\%$ vs $5.28\%$) due to the (\hov) model relative
to the (\cf) model (Wilcoxon signed-rank test: $W=17, p=0.026$) for the English phrases. A (\hov)-induced WER reduction was similarly seen for the OOV strings as well ($12.40\%$ vs. $15.57\%$), but the difference was not
statistically significant (Wilcoxon signed-rank test: $W=43.5, p=0.349$).

\subsubsection{Survey Results}

The post-block surveys asked the participants to rate their satisfaction
on a five-point Likert scale. The results from this survey question
showed that \textbf{the study participants, blinded from the conditions, assigned a significantly
higher satisfaction score to the (\hov) typing condition than the (\cf)
typing condition} (mean $4.0000 \pm 0.4830$ vs. $3.6250 \pm 0.6455$, paired t-test: $t(15)=3.223, p=0.006$).
Moreover, the surveys asked how difficult it was to type English phrases and OOV strings in the block.
Using a five-point Likert scale where 1 means ``very difficult'' and 5 means ``very easy'', the average scores of (\hov) and (\cf) were 3.875 and 3.375, respectively, for English phrases, and 2.875 and 2.813, respectively, for OOV strings.
The difference was significant for English phrases (paired t-test: $t(15)=4.899, p=1.93\times10^{-4}$) but not for OOV strings (paired t-test: $t(15) = 0.333, p=0.744$).
These were consistent with the significance test results of the typing metrics (WPM and WER) reported earlier. 

\subsubsection{Post Hoc Analysis} \label{subsec:online-posthoc}

\begin{figure}[t]
  \centering
  \includegraphics[width=0.95\linewidth]{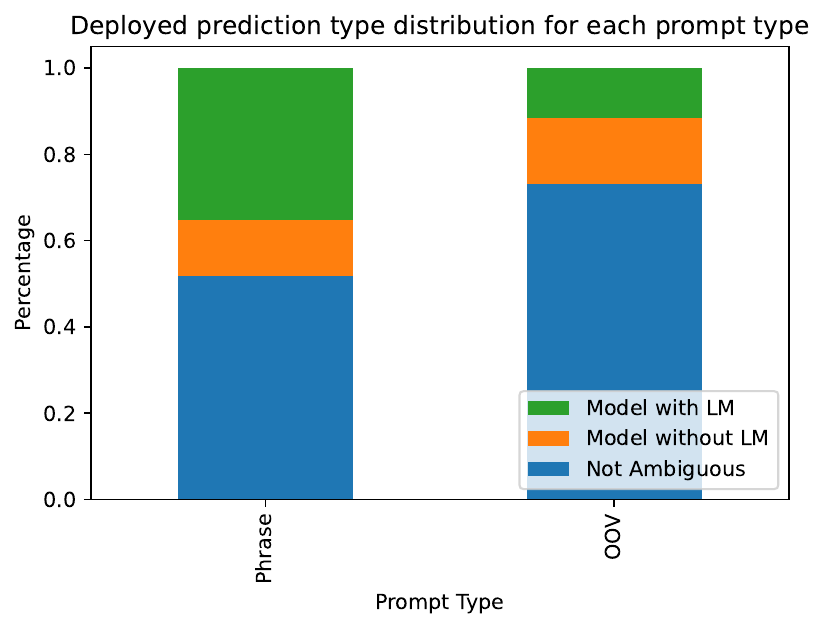}
  \caption{Prediction type distribution of touch points during deployment for each prompt type. 
  ``Not ambiguous'' touch points landed close enough to a key center and were not processed by the logistic-regression spatial models. 
  ``Model without LM'' means the spatial models were used but the LM scores were not available or were not informative (e.g., being equal for all the candidate keys).
  ``Model with LM'' means the spatial model scores were combined with the informative LM scores to make the final prediction. }  \label{fig:predtypes}
  \Description{A stacked bar chart showing the ratio of touch points predicted by the model with LM, the model without LM, and not predicted by the model because they are not ambiguous.}
\end{figure}

\begin{table}[t]
  \caption{Character error rates (CERs) $\downarrow \pm$ standard deviation (\%) of both models for each prompt type and LM setting.}
  \label{tab:cer}
  \setlength\tabcolsep{4.1pt}
  \begin{tabular}{L{0.9cm} R{1.5cm}  R{1.5cm} R{1.5cm} R{1.5cm}}
    \toprule
    Model & \multicolumn{2}{c}{English phrases} &  \multicolumn{2}{c}{OOV strings}\\
    & LM off & LM on  & LM off & LM on \\
    \midrule
    \multicolumn{5}{l}{\textit{Only touch points processed by the model}}\\
    (\cf) & 10.69 $\pm$ 4.98 & 4.76 $\pm$ 2.29 & 11.76 $\pm$ 7.23 & 28.89 $\pm$ 5.48\\
    (\hov) & 9.81 $\pm$ 5.33 & 3.66 $\pm$ 2.55 & 7.85 $\pm$ 5.70 & 22.84 $\pm$ 7.56\\
    \midrule
    \multicolumn{5}{l}{\textit{All touch points}}\\
    (\cf) & 6.04 $\pm$ 4.35 & 3.02 $\pm$ 2.05 & 3.55 $\pm$ 3.03 & 7.78 $\pm$ 3.23\\
    (\hov) & 5.72 $\pm$ 4.13 & 2.59 $\pm$ 2.08 & 2.91 $\pm$ 2.69 & 6.81 $\pm$ 4.34\\
    \bottomrule
\end{tabular}
\end{table}

The results so far indicate that for the English phrases used in this lab study, objective and
subjective measurements both support that the
touch heatmap-based decoding algorithm is not only capable of
supporting effective mobile tap typing but also significantly exceeds the conventional centroid-based
algorithm in accuracy, speed, and user experience. 
As for the OOV strings, increases in accuracy, speed, 
and user experience were also observed on average when typing with heatmap support than without. But 
these differences were smaller compared to the English phrases and did not
reach statistical significance under our sample size ($N=16$). To understand the reason for this difference, we analyzed the composition of taps across the two prompt types. The result, as Fig.~\ref{fig:predtypes} shows, indicates that the users landed their tap centroids in unambiguous on-key regions of the keyboard more often when entering the OOV strings than the English phrases ($73.2\%$ vs $51.8\%$), which likely resulted from the users' effort to accurately enter these unfamiliar prompts. If comparisons were limited to the portion of the taps that were ambiguous (i.e., the green and orange parts of the bars in Fig.~\ref{fig:predtypes}), though, significant reductions in CER were indeed seen under the heatmap (\hov) model than the (\cf) model ($22.84 \pm 7.32$ vs $28.89 \pm 5.30$, paired t-test: $t(15)=3.588, p=0.003$), again indicating the efficacy of the heatmaps input in improving the accuracy of keyboard decoding.

Additionally, we investigated the effect of the LM on these two prompt types.
During the live study, the LM was turned on together with the other additional techniques (i.e., skipping unambiguous cases and neighbor key filtering).
Here, we tried switching the LM on and off for the post hoc analysis (while always keeping the other additional techniques on) and observing the CER differences.
According to Table~\ref{tab:cer}, as expected, enabling the LM reduced the CER for English phrases by more than a half.
In contrast, it increased the CER for OOV strings for more than 100\%.
The result suggests that if the keyboard "knows" that the user is typing an OOV string, it should disable the LM.
This is a standard practice for password fields, for example.
However, generally, the user may type an OOV string (e.g., a foreign name or a novel acronym) amongst in-distribution English texts.
In this case, the user enters the OOV string while the LM is on, and that is when we can clearly see the benefit of the (\hov) model over the (\cf) model (CERs: $6.81\pm 4.20$ vs $7.78 \pm 3.13$ considering all touch points or $22.84 \pm 7.32$ vs $28.89 \pm 5.30$, considering only touch points processed by the model). 
Note that while LM is helpful for key decoding, our spatial modeling improvements complement the LM and benefit overall accuracy, even when the LM used is faced with OOV words. This also motivates the use of touch heatmaps in other touch-based applications, such as pin pads and calculators, where LM is inapplicable.

\section{Discussions and Future Work} \label{sec:discussions}

Our modeling study and user study results demonstrated that, relative to the traditional touch centroid data, touch heatmaps carry additional spatial information that can be used to reduce the error rate of decoding tapped keys during text entry on a smartphone.
Overall, our findings illuminate a direction of future research and engineering efforts aimed at enhancing the practical accuracy and usability of mobile typing.

\subsection{Heatmaps for Key Decoding}
\begin{table*}[t]
  \caption{Examples of user taps from Study 3 and the model predictions without using the language model. Each tap is aligned with the underlined character in the prompt. Despite the very close centroids, the ground truth labels of the two cases are different (\texttt{e} vs \texttt{r}). Only the models leveraging touch heatmap predicted both of them correctly.}
  \label{tab:examplepair}
  \begin{tabular}{L{2cm} C{4.5cm} L{2cm} C{4.5cm} L{2cm}}
    \toprule
    Prompt & \multicolumn{2}{c}{\texttt{Presidents driv\underline{e} expensive cars.}} & \multicolumn{2}{c}{\texttt{One hou\underline{r} is allotted for questions.}}\\
    \midrule
    Touch centroid and heatmap 
    &
    \multicolumn{2}{m{6.5cm}}{\includegraphics[width=6.5cm]{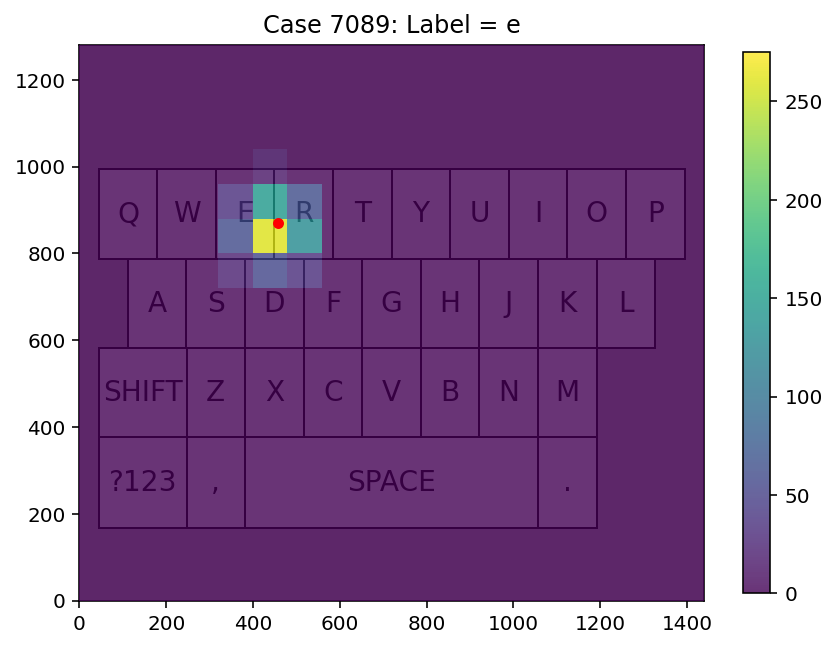}} & \multicolumn{2}{m{6.5cm}}{\includegraphics[width=6.5cm]{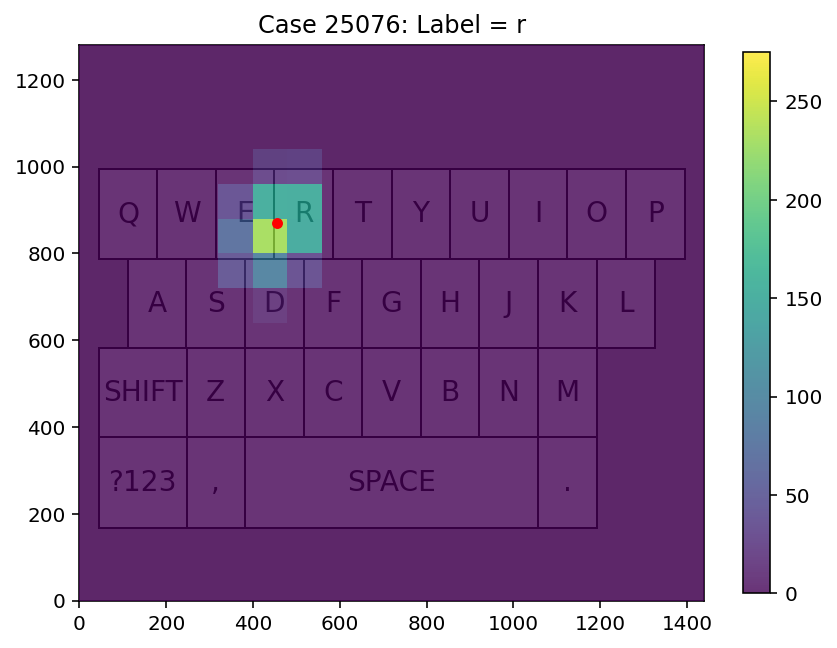}} \\
    
    Centroid position (x, y) & \multicolumn{2}{c}{(457, 152)} & \multicolumn{2}{c}{(456, 151)} \\
    \midrule
    \multicolumn{5}{l}{\textit{Predictions}} \\
    \quad On-key & \qquad\qquad\qquad \texttt{r} & \xmark & \qquad\qquad\qquad \texttt{r} & \cmark\\
    \quad (\cf) & \qquad\qquad\qquad \texttt{e}\quad (Prob. = 0.5197) & \cmark &  \qquad\qquad\qquad \texttt{e}\quad (Prob. = 0.5302) & \xmark\\
    \quad (\hov) & \qquad\qquad\qquad \texttt{e}\quad (Prob. = 0.6259) & \cmark &  \qquad\qquad\qquad \texttt{r}\quad (Prob. = 0.7474) & \cmark\\
    \quad (\chov) & \qquad\qquad\qquad \texttt{e}\quad (Prob. = 0.5915) & \cmark &  \qquad\qquad\qquad \texttt{r}\quad (Prob. = 0.7322) & \cmark\\
    \bottomrule
\end{tabular}
\end{table*}

Compared to touch centroids, touch heatmaps are at a lower level closer to the raw capacitive sensor input. When traditional mobile firmware postprocesses the touch sensor data into centroids, it approximates touches as point-like events, an assumption that is further justified by the economy and simplicity of the data format. Prior to this study, it was unclear whether this postprocessing discards information in touch images that has significant value for decoding touch events during typing, despite the knowledge that this information is useful for other purposes in mobile UIs \cite{boceck2019force, guarneri2013pca, le2018palmtouch, mayer2017estimating, quinn2021deep}. The results of the present study provide an affirmative answer to this question. The nature of such additional information originates from the fact that the contact between a user's typing digit (thumb or finger) and the touchscreen is neither point-like nor rigid, but instead reflects the geometric and elastic properties of the skin and underlying tissue. 

As Figs.~\ref{fig:misclassified-top}-\ref{fig:misclassified-bottom} show, the heatmaps are particularly informative when touch centroids land near the border between keys on the soft keyboard. In such cases, the relative capacitive inputs from the nearby touch sensors provide significant disambiguating power, especially in the lack of an accurate language model.
As a concrete example, Table~\ref{tab:examplepair} shows a pair of taps from Study 3 the centroids of which are very close to each other, i.e., (x, y) = (457, 152) vs (456, 151) at the pixel level. These taps are considered ambiguous because their centroids stay near the boundary between the keys \texttt{e} and \texttt{r}.
Due to the proximity of the two centroids, the (\cf) model predicted the same key (\texttt{e}) for both cases with the predicted probabilities approaching 0.5 (for they are near the key boundary).
However, based on the alignment with the prompts, only the first tap was in fact \texttt{e} while the true label of the second tap was \texttt{r}.
Looking at their heatmap images, we noticed an interesting difference that is the heatmap cells on the key \texttt{r} were activated more in the second tap than the first tap.  
Being aware of this information, the heatmap models, e.g., (\hov) and (\chov), therefore predicted both cases correctly with relatively higher predicted probabilities.
In other words, the results in this paper demonstrate that the touch heatmap is a better representation and treatment of the fat finger errors \cite{holz2011understanding,sivek2022spatial}, a prominent type of errors in mobile typing.
We further believe that, for smaller screens (e.g., smartwatches \cite{gordon2016watchwriter}), keyboards with smaller screen real estates, and those with smaller key sizes, the heatmaps may prove to be even more useful and merit further studies.

As an outlook, another factor that may affect the shape of the contact area, and hence the touch heatmap, is the position where the typing finger was at before moving to do the current tap. For example, with the QWERTY keyboard layout, the touch heatmaps of \texttt{s} in \texttt{cat\underline{s}} and \texttt{ha\underline{s}} could be different. In the former case, the user needs to move their finger from \texttt{t} to \texttt{s} ($\swarrow$) which can cause a different heatmap shape from the move from \texttt{a} to \texttt{s} ($\rightarrow$) in the latter case.
The current models treat each touch point individually, so they become agnostic on this information.
One interesting future work is, therefore, modifying the models to take information from the previous touch point(s) as additional input features. 
Another potential future work is to leverage heatmap frames beyond the first frame. For example, using the first $N$ frames of the target tap might further enhance the accuracy (due to the dynamic biomechanical information contained in the sequence). The optimal value of $N$ needs to be determined to best balance the trade-off between accuracy improvement and increases in latency and computational cost.
To support these studies, the bigram character distribution should be considered when constructing the prompt set for data collection, in addition to the unigram character distribution as inspected in Fig.~\ref{fig:unigram_distribution}.
In addition, a comprehensive study of alternative machine learning methods (apart from logistic regression) may lead to more optimized key decoding accuracy. 

\subsection{Generalization} 
The results from our offline test (Section~\ref{sec:ood}) and online test (Section~\ref{sec:userstudy}) showed that the ML model trained to decode touch heatmaps generalized satisfactorily to unseen users and unseen text, even in cases where the text was totally random and OOV, with decoding accuracies that consistently exceed the centroid-only baselines.
These findings, while informative, did not cover other aspects of generalizability that are also practically important for mobile text entry. The current study was limited to two-thumb typing, which tends to be the most common and fastest \cite{azenkot2012touch,palin2019people,reyal2015performance}, but not the only, posture used for mobile touchscreen typing. The decoding algorithms and models developed in this study will need to be further adapted for use with multiple hand postures. However, this would also create both challenges and opportunities for more accurate decoding, starting with effective and scalable heatmap data collection with different hand postures.

Apart from typing posture, in this paper, the models that utilized heatmaps are trained and evaluated for the same device model (Google Pixel 6 Pro), on the same keyboard size, and using the same screen size with the default orientation (portrait). In addition, all the 28 candidate keys are on the primary keyboard layout only (i.e., no number or punctuation except periods). Therefore, the generalizability of the learned models to other combinations of these factors is an unanswered question. Since the users' touch locations follow the size, location, and detailed key layout of the soft keyboard, it is unlikely that the learned ML models can generalize perfectly without adaptation if those factors are altered.
Effective generalization may call for remapping of the touch heatmap onto a different keyboard layout through image interpolation or warping, the efficacy of which is an open question for future studies. 
However, based on our small survey ($N = 46$) conducted to learn how people change the keyboard layout on their personal smartphones, we found that 30 people (65.2\%) have never changed the layout since they started using the phone whereas 13 people (28.3\%) said they have changed it only once. Only 3 people (6.5\%) have changed the layouts multiple times. This survey result shows that the default keyboard layout and size probably account for a majority of the user base, which ameliorates the concern about the burden of collecting heatmap data for numerous keyboard layouts.
Furthermore, models tailored to specific keyboard layouts and typing conditions can be trained at scale using federated learning. This allows multiple devices to collaboratively train the models without sharing user typing data, thereby preserving user privacy \cite{hard2018federated,zhang2023private}.

\section{Conclusion} \label{sec:conclusion}
This paper studies whether capacitive touch images from mobile touchscreens can enhance the accuracy of spatial models for decoding tap typing.
To do so, we trained and compared logistic regression-based spatial models with and without touch heatmaps as input features under three settings -- a leave-one-out cross-validation, an offline test, and an online test (i.e., a user study with the heatmap-based model deployed).
The results show that using or incorporating the touch heatmap indeed improved the decoding accuracy over using solely the touch centroid.
With the heatmap overlap vector representation, the user-study participants gave a significantly higher satisfaction score to the heatmap model compared to the centroid model on average, indicating that the benefit of touch heatmaps is perceivable to the users.
While this paper has shown that touch heatmaps contain rich and useful information for key decoding, the integration of heatmaps to the production keyboard system, which needs to support various keyboard layouts and typing environments, remains to be studied in the future.

\begin{acks}
We would like to thank Yuanbo Zhang, Gary Sivek, Philip Quinn, Michael Xuelin Huang, Cutter Coryell, Laurent Denoue, Fuxiao Xin, Jingtao Wang, and many other Gboard team members for their helpful input, comments, and discussions. We also extend our thanks to Sarika Amte and the Google UX research colleagues for their assistance in coordinating the user studies.
\end{acks}

\bibliographystyle{ACM-Reference-Format}
\bibliography{bibliography}

\newpage

\appendix

\section{Effect sizes} \label{app:effectsizes}
Tables~\ref{tab:cohen-id}-\ref{tab:cohen-cer} report the Cohen's $d$ effect sizes \cite{cohen2013statistical} of each result table in Sections~\ref{sec:modelingstudy}-\ref{sec:userstudy} in order to help interpret how impactful the different conditions were. In the following tables, \cf, \hf, and \hov denote the touch centroid, the flattened heatmap, and the heatmap overlap vector input features, respectively, while \chf is \cf and \hf concatenated and \chov is \cf and \hov concatenated.

\begin{table}[h]
  \caption{Cohen's $d$ effect size of method pairs in Table~\ref{tab:id} (Study 1) when using the spatial model (SM) only and when the additional techniques were incorporated.}
  \label{tab:cohen-id}
\setlength{\tabcolsep}{3.5pt}
\begin{tabular}{L{1.2cm} R{1cm} R{0.7cm} R{0.7cm} R{0.7cm} R{0.8cm} R{0.7cm} R{0.8cm}}
    \toprule
    Method & \multicolumn{7}{c}{Cohen's $d$ effect size $\uparrow$}\\ 
    & On-key & Dist. & (\cf) & (\hf) & (\chf) & (\hov) & (\chov)\\
    \midrule
    \multicolumn{8}{l}{\textit{SM only}}\\ 
    On-key & N/A & -1.04 & -1.73 & -2.30 & -2.25 & -2.22 &  -2.43 \\
    Distance & 1.04 & N/A & -1.62 & -2.16 & -2.11 & -2.09 & -2.28 \\
    (\cf) & 1.73 & 1.62 & N/A & -1.38 & -1.32 & -1.25 & -1.52 \\
    (\hf) & 2.30 & 2.16 & 1.38 & N/A & 0.07 & 0.92 & 0.85 \\
    (\chf) & 2.25 & 2.11 & 1.32 & -0.07 & N/A &0.85 & 0.78 \\
    (\hov) & 2.22 & 2.09 & 1.25 & -0.92 & -0.85 & N/A & -0.76 \\
    (\chov) & 2.43 & 2.28 & 1.52 & -0.85 & -0.78 & 0.76 & N/A \\
    \midrule
    \multicolumn{8}{l}{\textit{+ Additional techniques}}\\ 
    Distance & N/A & N/A & 1.13 & -1.54 & -1.49 & -0.84 & -1.06\\
    (\cf) &  N/A & -1.13 & N/A &-3.25 & -3.07 & -2.53 & -2.79 \\
    (\hf) &  N/A & 1.54 & 3.25 & N/A & 0.32 & 1.41 & 1.42 \\
    (\chf) &  N/A & 1.49 & 3.07 & -0.32 & N/A & 1.21 & 1.23\\
    (\hov) &  N/A & 0.84 & 2.53 & -1.41 & -1.21 & N/A & -0.38 \\
    (\chov) &  N/A & 1.06 & 2.79 & -1.42 & -1.23 & 0.38 & N/A\\
    \bottomrule
\end{tabular}

\end{table}

\begin{table}[h]
  \caption{Cohen's $d$ effect size of method pairs in Table~\ref{tab:ood} (Top) (Study 2: OfflineTest-Subset) when using the spatial model (SM) only and when the additional techniques were incorporated.}
  \label{tab:cohen-ood-subset}
\setlength{\tabcolsep}{3.5pt}
\begin{tabular}{L{1.2cm} R{1cm} R{0.7cm} R{0.7cm} R{0.7cm} R{0.8cm} R{0.7cm} R{0.8cm}}
    \toprule
    Method & \multicolumn{7}{c}{Cohen's $d$ effect size $\uparrow$}\\ 
    & On-key & Dist. & (\cf) & (\hf) & (\chf) & (\hov) & (\chov)\\
    \midrule
    \multicolumn{8}{l}{\textit{SM only}}\\ 
    On-key & N/A	&	-0.71	&	-0.18	&	-0.36	&	-0.41	&	-0.42	&	-0.57	\\
Distance & 0.71	&	N/A	&	-0.10	&	-0.27	&	-0.32	&	-0.33	&	-0.47	\\
(\cf) &  0.18	&	0.10	&	N/A	&	-0.49	&	-0.68	&	-0.32	&	-0.58	\\
(\hf) & 0.36	&	0.27	&	0.49	&	N/A	&	-0.39	&	-0.01	&	-0.21	\\
(\chf) &  0.41	&	0.32	&	0.68	&	0.39	&	N/A	&	0.08	&	-0.11	\\
(\hov) &  0.42	&	0.33	&	0.32	&	0.01	&	-0.08	&	N/A	&	-0.25	\\
(\chov) & 0.57	&	0.47	&	0.58	&	0.21	&	0.11	&	0.25	&	N/A	\\
    \midrule
    \multicolumn{8}{l}{\textit{+ Additional techniques}}\\ 
    Distance & N/A	&	N/A	&	1.10	&	-0.08	&	0.01	&	-0.03	&	-0.09	\\
(\cf) &  N/A	&	-1.10	&	N/A	&	-1.62	&	-1.53	&	-1.66	&	-2.34	\\
(\hf) &  N/A	&	0.08	&	1.62	&	N/A	&	0.94	&	0.10	&	0.01	\\
(\chf) & N/A	&	-0.01	&	1.53	&	-0.94	&	N/A	&	-0.06	&	-0.17	\\
(\hov) & N/A	&	0.03	&	1.66	&	-0.10	&	0.06	&	N/A	&	-0.21	\\
(\chov) & N/A	&	0.09	&	2.34	&	-0.01	&	0.17	&	0.21	&	N/A	\\
    \bottomrule
\end{tabular}

\end{table}

\newpage

\begin{table}[t]
  \caption{Cohen's $d$ effect size of method pairs in Table~\ref{tab:ood} (Bottom) (Study 2: OfflineTest-Unseen) when using the spatial model (SM) only and when the additional techniques were incorporated.}
  \label{tab:cohen-ood-unseen}
\setlength{\tabcolsep}{3.5pt}
\begin{tabular}{L{1.2cm} R{1cm} R{0.7cm} R{0.7cm} R{0.7cm} R{0.8cm} R{0.7cm} R{0.8cm}}
    \toprule
    Method & \multicolumn{7}{c}{Cohen's $d$ effect size $\uparrow$}\\ 
    & On-key & Dist. & (\cf) & (\hf) & (\chf) & (\hov) & (\chov)\\
    \midrule
    \multicolumn{8}{l}{\textit{SM only}}\\ 
On-key & N/A	&	0.15	&	-0.51	&	-0.68	&	-0.70	&	-1.00	&	-0.86	\\
Distance & -0.15	&	N/A	&	-0.56	&	-0.74	&	-0.77	&	-1.05	&	-0.92	\\
(\cf) & 0.51	&	0.56	&	N/A	&	-0.22	&	-0.35	&	-0.72	&	-0.68	\\
(\hf) & 0.68	&	0.74	&	0.22	&	N/A	&	-0.56	&	-0.90	&	-0.84	\\
(\chf) & 0.70	&	0.77	&	0.35	&	0.56	&	N/A	&	-0.75	&	-0.71	\\
(\hov) & 1.00	&	1.05	&	0.72	&	0.90	&	0.75	&	N/A	&	0.44	\\
(\chov) & 0.86	&	0.92	&	0.68	&	0.84	&	0.71	&	-0.44	&	N/A	\\
    \midrule
    \multicolumn{8}{l}{\textit{+ Additional techniques}}\\ 
Distance & N/A	&	N/A	&	0.42	&	-0.55	&	-0.54	&	-0.58	&	-0.48	\\
(\cf) & N/A	&	-0.42	&	N/A	&	-1.25	&	-1.27	&	-1.89	&	-1.58	\\
(\hf) & N/A	&	0.55	&	1.25	&	N/A	&	0.21	&	-0.36	&	-0.15	\\
(\chf) & N/A	&	0.54	&	1.27	&	-0.21	&	N/A	&	-0.42	&	-0.21	\\
(\hov) & N/A	&	0.58	&	1.89	&	0.36	&	0.42	&	N/A	&	0.24	\\
(\chov) & N/A	&	0.48	&	1.58	&	0.15	&	0.21	&	-0.24	&	N/A	\\
    \bottomrule
\end{tabular}

\end{table}

\begin{table}[t]
  \caption{Absolute Cohen's $d$ effect size between (\cf) and (\hov) of typing metrics in Table~\ref{tab:metrics} (Study 3). Note that (\hov) outperformed (\cf) in both metrics and both prompt types.}
  \label{tab:cohen-metrics}
\setlength{\tabcolsep}{3.5pt}
\begin{tabular}{R{1.8cm} R{1.8cm} R{1.8cm} R{1.8cm}}
    \toprule
    \multicolumn{4}{c}{Absolute Cohen's $d$ effect size}\\
    WPM-Phrase & WPM-OOV  & WER-Phrase & WER-OOV \\
    \midrule
    0.6197 & 0.1775 & 0.5129 & 0.2942 \\ 
    \bottomrule
\end{tabular}

\end{table}

\begin{table}[t]
  \caption{Absolute Cohen's $d$ effect size between (\cf) and (\hov) of character error rates (CERs) in Table~\ref{tab:cer} (Study 3). Note that (\hov) outperformed (\cf) in both English phrases and OOV strings regardless of whether the LM is on.}
  \label{tab:cohen-cer}
\setlength{\tabcolsep}{3.5pt}
  \begin{tabular}{R{1.8cm} R{1.8cm} R{1.8cm} R{1.8cm}}
    \toprule
    \multicolumn{4}{c}{Absolute Cohen's $d$ effect size}\\
    Phrase-LM off & Phrase-LM on  & OOV-LM off & OOV-LM on \\
    \midrule
    \multicolumn{4}{l}{\textit{Only touch points processed by the model}}\\
    0.2819 & 1.0725 & 0.4785 & 0.8970 \\
    \midrule
    \multicolumn{4}{l}{\textit{All touch points}} \\
    0.1598 & 0.7319 & 0.2527 & 0.4009 \\
    \bottomrule
\end{tabular}

\end{table}

\newpage

\section{Additional Simulation for Study 3} \label{app:simulate-study3}
In Study 3, we used the (\hov) model as a representative of the heatmap-based models due to its best performance when decoding unseen prompts in Study 2. 
In this section, we simulated Study 3 for the other three heatmap models, (\hf), (\chf), and (\chov), on the same set of taps (\hov) processed. The CERs averaged across all participants are shown in Table~\ref{tab:simulate-study3}. 
Overall, the (\hov) model was still the best model among these feature sets. 

\begin{table*}[t]
  \caption{Character error rates (CERs) $\downarrow \pm$ standard deviation (\%) when applying the models on the same set of taps (\hov) processed in Study 3. The results are reported for each prompt type and LM setting.}
  \label{tab:simulate-study3}
  \setlength\tabcolsep{6pt}
  \begin{tabular}{L{1.2cm} R{1.5cm}  R{1.5cm} R{1.5cm} R{1.5cm} R{1.5cm} R{1.5cm}}
    \toprule
    Model & \multicolumn{2}{c}{English phrases} &  \multicolumn{2}{c}{OOV strings} &  \multicolumn{2}{c}{Overall}\\
    & LM off & LM on  & LM off & LM on & LM off & LM on \\
    \midrule
    \multicolumn{7}{l}{\textit{Only touch points processed by the model}}\\
    (\cf) & 11.55 $\pm$ 5.47 & 5.00 $\pm$ 2.67 & 12.39 $\pm$ 6.51 & 27.79 $\pm$ 6.81 & 11.64 $\pm$ 5.42 & 6.63 $\pm$ 2.99 \\
    (\hf) & 10.86 $\pm$ 5.70 & 4.27 $\pm$ 2.94 & 7.22 $\pm$ 6.23 & 22.20 $\pm$ 7.35 & 10.63 $\pm$ 5.39 & 5.54 $\pm$ 3.06 \\
    (\chf) & 10.73 $\pm$ 5.82 & 4.19 $\pm$ 3.03 & 7.22 $\pm$ 6.23 & 22.61 $\pm$ 7.61 & 10.51 $\pm$ 5.48 & 5.49 $\pm$ 3.18 \\
    (\hov) & 9.80 $\pm$ 5.32 & 3.65 $\pm$ 2.53 & 7.85 $\pm$ 5.70 & 22.84 $\pm$ 7.56 & 9.68 $\pm$ 5.13 & 5.00 $\pm$ 2.87 \\
    (\chov) & 9.81 $\pm$ 5.30 & 3.76 $\pm$ 2.62 & 8.06 $\pm$ 5.31 & 22.94 $\pm$ 7.23 & 9.70 $\pm$ 5.07 & 5.11 $\pm$ 2.90 \\
    \midrule
    \multicolumn{7}{l}{\textit{All touch points}}\\
    (\cf) & 6.55 $\pm$ 4.41 & 3.21 $\pm$ 2.24 & 4.22 $\pm$ 3.90 & 8.25 $\pm$ 5.28 & 6.28 $\pm$ 4.23 & 3.84 $\pm$ 2.46 \\
    (\hf) & 6.27 $\pm$ 4.46 & 2.92 $\pm$ 2.39 & 2.68 $\pm$ 2.73 & 6.55 $\pm$ 4.12 & 5.83 $\pm$ 4.05 & 3.37 $\pm$ 2.39 \\
    (\chf) & 6.21 $\pm$ 4.57 & 2.89 $\pm$ 2.47 & 2.68 $\pm$ 2.73 & 6.69 $\pm$ 4.26 & 5.79 $\pm$ 4.14 & 3.36 $\pm$ 2.49 \\
    (\hov) & 5.72 $\pm$ 4.12 & 2.58 $\pm$ 2.07 & 2.91 $\pm$ 2.69 & 6.81 $\pm$ 4.34 & 5.38 $\pm$ 3.81 & 3.11 $\pm$ 2.21 \\
    (\chov) & 5.74 $\pm$ 4.15 & 2.65 $\pm$ 2.12 & 2.91 $\pm$ 2.63 & 6.81 $\pm$ 4.32 & 5.40 $\pm$ 3.82 & 3.17 $\pm$ 2.24 \\
    \bottomrule
\end{tabular}
\end{table*}

\section{Training and Validation Losses} \label{app:losses}
To demonstrate reliability of the models, we report the training and validation cross-entropy losses in Table~\ref{tab:losses}. For Study 1, the losses were averaged across the 24 folds of the leave-one-out cross validation. Meanwhile, for Study 3, the losses were reported for the single participant split used to train the models in the study. We do not report the losses for Study 2 since it used the same set of models as Study 1.

\begin{table}[h]
  \caption{Training and validation cross-entropy losses of the models. For Study 1, we report both the average losses and the standard deviation computed across the 24 folds of the leave-one-out cross validation.}
  \label{tab:losses}
  \setlength\tabcolsep{4.1pt}
  \begin{tabular}{L{0.9cm} R{1.7cm}  R{1.8cm} R{1.2cm} R{1.4cm}}
    \toprule
    Model & \multicolumn{2}{c}{Study 1} &  \multicolumn{2}{c}{Study 3}\\
    & Training & Validation  & Training & Validation \\
    \midrule
    (\cf) & 0.226 $\pm$ 0.004 & 0.229 $\pm$ 0.018 & 0.246 & 0.255 \\
    (\hf) & 0.102 $\pm$ 0.005 & 0.116 $\pm$ 0.024 & 0.100 & 0.120 \\
    (\chf) & 0.098 $\pm$ 0.004 & 0.110 $\pm$ 0.023 & 0.095 & 0.114 \\
    (\hov) & 0.141 $\pm$ 0.007 & 0.146 $\pm$ 0.027 & 0.135 & 0.157 \\
    (\chov) & 0.124 $\pm$ 0.006 & 0.128 $\pm$ 0.025 & 0.118 & 0.136 \\
    \bottomrule
\end{tabular}
\end{table}

\end{document}